\documentclass[fleqn,usenatbib]{mnras}

\usepackage{newtxtext,newtxmath}
\usepackage[T1]{fontenc}
\usepackage{ae,aecompl}

\usepackage{graphicx}	% Including figure files
\usepackage{amsmath}	% Advanced maths commands
\usepackage{amssymb}	% Extra maths symbols
\usepackage{booktabs}
\usepackage{multirow}
\usepackage{mathtools}
\usepackage{silence}
\DeclareMathOperator\arcsinh{arcsinh}

\title[Green Valley Structure]{Galaxy and Mass Assembly (GAMA): Variation in Galaxy Structure Across the Green Valley}
\author[L.~S.~Kelvin et al.]{
\parbox{\textwidth}{
\raggedright
Lee~S.~Kelvin,$^{1}$
Malcolm~N.~Bremer,$^{2}$
Steven~Phillipps,$^{2}$
Philip~A.~James,$^{1}$
Luke~J.~M.~Davies,$^{3}$
Roberto~De~Propris,$^{4}$
Amanda~J.~Moffett,$^{5}$
Susan~M.~Percival,$^{1}$
Ivan~K.~Baldry,$^{1}$
Chris~A.~Collins,$^{1}$
Mehmet~Alpaslan,$^{6}$
Joss~Bland-Hawthorn,$^{7}$
Sarah~Brough,$^{8}$
Michelle~Cluver,$^{9}$
Simon~P.~Driver,$^{3,10}$
Abdolhosein~Hashemizadeh,$^{3}$
Benne~W.~Holwerda,$^{11}$
Jarkko~Laine,$^{12}$
Maritza~A.~Lara-Lopez,$^{13}$
Jochen~Liske,$^{12}$
Witold~Maciejewski,$^{1}$
Nicola~R.~Napolitano,$^{14}$
Samantha~J.~Penny,$^{15}$
Cristina~C.~Popescu,$^{16,17}$
Anne~E.~Sansom,$^{16}$
Will~Sutherland,$^{18}$
Edward~N.~Taylor,$^{19}$
Eelco~van~Kampen$^{20}$
and Lingyu~Wang$^{21,22}$
}\vspace{0.5cm}\\
\parbox{\textwidth}{
$^{1}$Astrophysics Research Institute, Liverpool John Moores University, IC2, Liverpool Science Park, 146 Brownlow Hill, Liverpool, L3 5RF, UK\\
$^{2}$Astrophysics Group, H.H. Wills Physics Laboratory, University of Bristol, Tyndall Avenue, Bristol, BS8 1TL, UK\\
$^{3}$International Center for Radio Astronomy Research, The University of Western Australia, 35 Stirling Highway, Crawley, WA 6009, Australia\\
$^{4}$Finnish Centre for Astronomy with ESO, University of Turku, V{\"a}is{\"a}al{\"a}ntie 20, Piikki{\"o}, Finland\\
$^{5}$Department of Physics \& Astronomy, Vanderbilt University, Nashville TN 37240 USA\\
$^{6}$Center for Cosmology and Particle Physics, Department of Physics, New York University, 726 Broadway, New York, NY 10003, USA\\
$^{7}$Sydney Institute for Astronomy, School of Physics A28, University of Sydney, NSW 2006, Australia\\
$^{8}$School of Physics, University of New South Wales, NSW 2052, Australia\\
$^{9}$Department of Physics and Astronomy, University of the Western Cape, Robert Sobukwe Road, Bellville, 7535, South Africa\\
$^{10}$School of Physics and Astronomy, University of St Andrews, North Haugh, St Andrews KY16 9SS, UK\\
$^{11}$Department of Physics and Astronomy, 102 Natural Science Building, University of Louisville, Louisville KY 40292, USA\\
$^{12}$Hamburger Sternwarte, Universit\"{a}t Hamburg, Gojenbergsweg 112, 21029 Hamburg, Germany\\
$^{13}$Dark Cosmology Centre, Niels Bohr Institute, University of Copenhagen, Juliane Maries Vej 30, DK-2100 Copenhagen, Denmark\\
$^{14}$INAF -- Osservatorio Astronomico di Capodimonte, Salita Moiariello, 16, I-80131 Napoli, Italy\\
$^{15}$Institute of Cosmology and Gravitation, University of Portsmouth, Burnaby Road, Portsmouth PO1 3FX, UK\\
$^{16}$Jeremiah Horrocks Institute, School of Physical Sciences and Computing, University of Central Lancashire, Preston, PR1 2HE, UK\\
$^{17}$The Astronomical Institute of the Romanian Academy, Str. Cutitul de Argint 5, Bucharest, Romania\\
$^{18}$School of Physics and Astronomy, Queen Mary University of London, Mile End Road, London E1 4NS, UK\\
$^{19}$Centre for Astrophysics and Supercomputing, Swinburne University of Technology, Hawthorn 3122, Australia\\
$^{20}$ESO, Karl-Schwarzschild-Str. 2, 85748 Garching bei Muenchen, Germany\\
$^{21}$SRON Netherlands Institute for Space Research, Landleven 12, 9747 AD, Groningen, The Netherlands\\
$^{22}$Kapteyn Astronomical Institute, University of Groningen, Postbus 800, 9700 AV, Groningen, The Netherlands\\
}
\vspace{-0.75cm}
}
\date{Accepted XXX. Received YYY; in original form ZZZ}
\pubyear{2018}

\begin{document}
\label{firstpage}
\pagerange{\pageref{firstpage}--\pageref{lastpage}}
\maketitle

%%%%%%%%%%%%%%%%%%%%%%%%%%%%%%%%%%%%%%%%%%%%%%%%%%

%%%%%%%%%%%%%%%%%%%% ABSTRACT %%%%%%%%%%%%%%%%%%%%

\clearpage

\begin{abstract}
Using a sample of $472$ local Universe ($z<0.06$) galaxies in the stellar mass range $10.25<\log\mathcal{M}_{\star}/\mathcal{M}_{\odot}<10.75$, we explore the variation in galaxy structure as a function of morphology and galaxy colour. Our sample of galaxies is sub-divided into red, green and blue colour groups and into elliptical and non-elliptical (disk-type) morphologies. Using KiDS and VIKING derived postage stamp images, a group of eight volunteers visually classified bars, rings, morphological lenses, tidal streams, shells and signs of merger activity for all systems. We find a significant surplus of rings ($2.3\sigma$) and lenses ($2.9\sigma$) in disk-type galaxies as they transition across the green valley. Combined, this implies a joint ring/lens green valley surplus significance of $3.3\sigma$ relative to equivalent disk-types within either the blue cloud or the red sequence. We recover a bar fraction of $\sim44\%$ which remains flat with colour, however, we find that the presence of a bar acts to modulate the incidence of rings and (to a lesser extent) lenses, with rings in barred disk-type galaxies more common by $\sim20-30$ percentage points relative to their unbarred counterparts, regardless of colour. Additionally, green valley disk-type galaxies with a bar exhibit a significant $3.0\sigma$ surplus of lenses relative to their blue/red analogues. The existence of such structures rules out violent transformative events as the primary end-of-life evolutionary mechanism, with a more passive scenario the favoured candidate for the majority of galaxies rapidly transitioning across the green valley.
\end{abstract}


%%%%%%%%%%%%%%%%%%%%%%%%%%%%%%%%%%%%%%%%%%%%%%%%%%

%%%%%%%%%%%%%%%%% BODY OF PAPER %%%%%%%%%%%%%%%%%%

\section{Introduction}
\label{sec:introduction}

The presence or absence of a disk plays a fundamental role in galaxy structure and dynamics. Non-disk galaxies generally have a smooth appearance and include Hubble's elliptical galaxy class. Those having a disk component are generally classified as spirals and S0s. Historically, ellipticals and S0s have been referred to as early-type galaxies whilst most spirals have been referred to as late-type galaxies. Early type elliptical and lenticular galaxies (ETGs) are typically red, quiescent and visually smooth systems, whilst late type spiral galaxies (LTGs) are typically blue, star-forming, potentially barred and visually complex systems \citep{Jeans1919,Reynolds1920,Hubble1926}. Beyond visual morphology, galaxy bimodality in various feature planes is well documented in the literature, for example, in colour--magnitude \citep{Tully1982,Baldry2004}, colour-colour \citep{Strateva2001}, size--magnitude \citep{Simard2011}, size--colour \citep{Kelvin2014a}, colour--concentration \citep{Driver2006,Kelvin2012}, environment--concentration \citep{Dressler1980,Hiemer2014}, star-formation rate (SFR)--concentration \citep{Schiminovich2007}, environment--stellar mass \citep{Baldry2006}, SFR--stellar mass \citep{Smethurst2015} and colour--stellar mass \citep{Kelvin2014b,Taylor2015}. In all of these planes, a representative volume-limited sample of galaxies will primarily cluster into two regimes loosely corresponding to the aforementioned late-type and early-type galaxies. In the colour--magnitude or colour--mass diagram (CMD), these groupings are denoted \textit{blue-cloud} and \textit{red-sequence}, respectively. The relatively underdense region between these two populations may simply be the overlapping tails of two Gaussian distributions which model the two populations \citep{Baldry2004,Taylor2015}. However, the so-called \textit{green valley} has been of particular interest in recent years as the region in CMD space populated by galaxies transitioning between the blue cloud and the red sequence \citep{Martin2007,Wyder2007}.

Owing to its relative under-density, the timescale for transition of galaxies across the green valley is thought to be relatively rapid ($\sim1-2\,\mathrm{Gyr}$; \citealp{Bremer2018}, hereafter Paper I; see also \citealp{Martin2007}). Possible mechanisms for rapid quenching from the blue cloud to the red sequence include major-merger induced starbursts \citep{Schiminovich2007}, AGN feedback \citep{Martin2007}, shock excitation / turbulent heating \citep{Cluver2013}, morphological quenching \citep{Martig2009} or simply running low on the cold gas necessary to fuel star formation as a result of the relatively old age of individual systems (Paper I) \citep[see also][]{Masters2010,Fang2012}. The latter scenario is in excellent agreement with a gradual evolutionary process consistent with a unitary population of galaxies as proposed by \citet{Eales2017} \citep[see also][]{Oemler2017,Feldmann2017b,Eales2018}. \citet{Schawinski2014} favour a two-mode approach to green valley transition dependent upon morphological type, with LTGs slowly exhausting their gas over several billion years and ETGs requiring an initial major merger followed by a subsequent morphological transition. A different, multi-mode approach to green valley transition has also recently been favoured by \citet{Rowlands2018}, with various evolutionary pathways correlating with galaxies of a given stellar mass and epoch. Whilst blue to red transitions are believed to be the dominant mode of travel in the CMD as evidenced by the preponderance of red galaxies at its most massive end, this is not the only pathway to the green valley. \citet{Thilker2010} show for NGC 404 that the merger of an ETG and a gas-rich dwarf may trigger the rejuvenation of a previously red sequence galaxy back into the green valley or beyond \citep[see also][]{Fang2012}, perhaps related to the low-redshift population of dusty early type galaxies found residing in the green valley \citep{Agius2013}. Similarly, a galaxy in the blue cloud may suffer from a temporary ($\ll1\,\mathrm{Gyr}$) cessation of star formation eliciting a brief foray into the green valley \citep{Feldmann2017a}. Whilst these red to blue mechanisms are not believed to account for a significant fraction of traffic across the green valley \citep[$\sim1\%$,][]{Trayford2016}, they are worth noting here.

The terms `early-type' and `late-type' do not correspond to the evolutionary pathway taken by a galaxy, nor were they ever intended to \citep{Sandage2005,Baldry2008b}. However, it is clear that over the course of its lifetime a galaxy will undergo a multitude of evolutionary processes with each leaving behind a distinct structural watermark on its resultant morphology \citep{Buta2013}. For example, the effects of minor merging and tidal interactions \citep{Park2008}, strangulation \citep{Larson1980,Kauffmann1993,White1993,Diaferio2001}, harassment \citep{Moore1996} and ram pressure stripping \citep{Gunn1972} are all now understood to be important mechanisms in modifying the visual morphology of a galaxy and causing the emergence of various galactic structures.

Beyond the ubiquitous spheroid, disk and bar, a number of additional morphological structures also exist in nature (see Section \ref{sec:results} for examples), with each indicating a prior evolutionary mechanism at play. Rings, large diffuse and relatively radially thin structures encircling the central galaxy region, were initially noted in observations performed by Lord Rosse at Birr Castle in the 1850s and 1860s (NGC 4725, for example), with the first outer ring discovered in NGC 1291 by \citet{Perrine1922}. The (r) and (s) nomenclature for identifying galaxies with and without a ring, respectively, was introduced by Allan Sandage in the Hubble Atlas, and further expanded upon by \citet{deVaucouleurs1959} \citep[see also][]{Sandage1961}. Outer rings and pseudorings, apparent rings formed from tightly wound spiral arms, are found within $\sim10\%-30\%$ of disk galaxies \citep{Buta1996}. Many of these structures match those predicted to form at the Outer Lindblad Resonance of the bar during the evolutionary progression of a disk \citep{Schwarz1981}, with some instead linked to the outer 4:1 resonance \citep{Buta2017a}. Unlike an outer ring, inner rings are most often found in barred galaxies with a similar radius to that of the bar \citep{deVaucouleurs1959,Freeman1975,Kormendy1979}, with the area swept out by the bar and interior to the inner ring comprising the `Star Formation Desert' \citep{James2015,James2016,James2018}. Inner rings and pseudorings are found within $\sim50\%$ of disk galaxies \citep{Buta1996} and believed to have their origins in the inner 4:1 ultraharmonic resonance \citep{Schwarz1984}. A morphological lens, such as that found within NGC 4909 \citep{Buta2007}, is an elliptical feature falling between the central bulge and the outer disk. It is typically characterised as a region of relatively constant surface brightness in the core before falling off sharply at some outer edge, hence the coinage \textit{plateau} to help differentiate a morphological lens from the multitude of other features in the realm of astronomy also labelled \textit{lens}. The origin of the lens remains unclear, however; \citet{Kormendy1979} ascribes these features to a dissolved or dissolving bar. Some lenses may have their origin in ring evolution, as there exists a lens analogue for each type of ring \citep[][and references therein]{Buta2015}. Alternatively, and quite distinct from a ring analogue lens, \citet{Salo2017} \citep[see also][]{Laurikainen2017} suggest that a vertically thick boxy/peanut section of the bar as viewed edge-on may manifest as a lens-like `barlens' feature when viewed face-on. Indeed, some form of dynamical link between the bar, lens and ring is evident in many systems, with the inner ring often connecting at the extreme radius of the bar, the outer ring being found at roughly twice the radius of the bar, and lens-like structures often exhibiting similar radial sizes to that of their equivalent ring counterparts \citep{Laurikainen2013,Comeron2014}. Additional types of ring include central $\sim1.5\,\mathrm{kpc}$ diameter nuclear rings associated with the bar \citep[see][and references therein]{Comeron2010,Knapen2010} and catastrophic rings, rings formed via a violent direct collision along the polar axis of the primary galaxy or via accretion of a satellite galaxy into an outer ring \citep[cf. Hoag's Object,][]{Arp1966,Schweizer1987,Appleton1996}. 

Gravitational interactions also lead to disturbance or the formation of galaxy structure. In the local Universe, \citet{Knapen2009} find that $2\%-4\%$ of bright ($M_B\lesssim-18$) galaxies are currently in the process of interacting or merging, leading to a disturbed or peculiar morphology. Shells (ripples), first noted around elliptical galaxies \citep{Malin1980,Malin1983}, are the remnants of a relatively recent dry minor merger with a lower mass disk galaxy \citep{Athanassoula1985,Schweizer1988,Kormendy1989}. Furthermore, \citet{Salo2000a,Salo2000b} show that both parabolic and bound encounters between galaxies leave behind tidal tails and streams of stellar material in their wake \citep[see also][and references therein]{Cullen2007,Martinez-Delgado2010,Martinez-Delgado2015}. 

In this paper we explore the variation of galaxy structure across the green valley with some of the deepest wide-area multi-wavelength imaging data currently available, and discuss implications this may have for our current understanding of green valley transitions. This paper is organized as follows: Section \ref{sec:data} summarises the survey data used as an input for this study, detailing our sample definition and the derivation of various galaxy properties. Section \ref{sec:census} describes our structural classification methodology (the \textit{Green Valley Census}). The results of this study are presented in Section \ref{sec:results} and a discussion of their implications in Section \ref{sec:discussion}. A standard cosmology of $H_0=70\,\mathrm{km}\,\mathrm{s}^{-1}\,\mathrm{Mpc}^{-1}$, $\Omega_M=0.3$, $\Omega_\Lambda=0.7$ is assumed throughout. Magnitudes are presented in the AB system.

\section{Data}
\label{sec:data}

\subsection{Surveys}
\label{sec:surveys}

The Galaxy And Mass Assembly survey \citep[GAMA\footnote{www.gama-survey.org},][]{Driver2009,Liske2015,Baldry2018} is a spectroscopic and multi-wavelength imaging campaign designed to study galaxy structure on scales of $1$ kpc to $1$ Mpc in the local ($z\lesssim0.5$) Universe. GAMA consists of $\sim$$300000$ galaxies down to a nominal apparent magnitude limit of $r=19.8$ mag over $286$ $\mathrm{deg}^{2}$ spread over $5$ patches of sky. Imaging has been collected and reprocessed from a number of wide area surveys including the Sloan Digital Sky Survey \citep[SDSS,][]{York2000,Abazajian2009}, the UK Infrared Deep Sky Survey Large Area Survey \citep[UKIDSS-LAS,][]{Lawrence2007,Dye2006,Warren2007a,Warren2007b}, the VLT Survey Telescope (VST) Kilo-Degree Survey \citep[KiDS,][]{Capaccioli2011,deJong2013}, and the Visible and Infrared Survey Telescope for Astronomy (VISTA) Kilo-Degree Infrared Galaxy Survey \citep[VIKING,][]{Arnaboldi2007,Edge2013,Sutherland2015}. See \cite{Driver2009} and references therein for further details. Owing to its excellent completeness and numerous value-added data catalogues, GAMA is an ideal survey for target selection and subsequent examination of galaxies in and around the green valley.

The $2.65\,\mathrm{m}$ VST \citep{Capaccioli2011} located at Paranal Observatory, Chile, has been specifically designed for optical wide-area imaging surveys. KiDS \citep[][]{deJong2013,deJong2015,deJong2017} is one such survey, initially aiming to cover $1500$ square degrees\footnote{The final imaged region may cover closer to $1350$ square degrees.} principally across two regions on the sky: KiDS North (KiDS-N) on the celestial equator, and KiDS South (KiDS-S) covering the South Galactic Pole. Additional smaller fields are targeted to provide maximal overlap with surveys such as GAMA and other contemporary imaging and redshift surveys. The OmegaCAM instrument \citep{Kuijken2011} consists of $8\times4=32$ CCDs, each with a pixel format of $2\mathrm{k}\times4\mathrm{k}$ and a per-pixel scale of $0.2$ arcseconds. The KiDS regions are observed with OmegaCAM in the Sloan $ugri$ passbands, with exposure times of $900$, $900$, $1800$ and $1080$ seconds and observed magnitude limits ($5\sigma$ in a $2$ arcsecond aperture) of $24.8$, $25.4$, $25.2$ and $24.2$ mag, respectively. Observations are conducted using an overlapping dither pattern ($5\times$ dithers in $gri$, $4\times$ dithers in $u$) in order to avoid holes in the contiguous survey region. The significant depth afforded by KiDS in the optical allows for the large-scale characterisation of galaxy structure which would otherwise be undetectable in shallower wide-area imaging surveys such as SDSS.

The $4.1\,\mathrm{m}$ VISTA \citep[][]{Sutherland2015}, also located at Paranal Observatory in Chile, is the near-infrared (NIR) counterpart to the VST. VIKING \citep[][]{Edge2013} is a $1350$ square degree NIR imaging campaign covering essentially the same sky as the aforementioned KiDS regions. VIKING images the sky in the $ZYJHK_s$ passbands using the VISTA IR Camera \citep[VIRCAM,][]{Dalton2010,Hummel2010}: a NIR imager covering $0.6\deg$ per pointing and consisting of $4\times4=16$ HgCdTe near-infrared detectors each of dimension $2\mathrm{k}\times2\mathrm{k}$. Median survey depths ($5\sigma$) in $ZYJHKs$ of $23.1$, $22.3$, $22.1$, $21.5$ and $21.2$ mag are provided across all imaged regions. NIR VIKING imaging benefits from relatively smaller extinction effects than as observed in the optical, and allows for the analysis of older stellar populations within galaxies alongside the younger stellar populations visible in the optical.

\subsection{Sample}
\label{sec:sample}

Our ultimate sample has been selected to largely match the selection of the higher redshift sample in use in Paper I, such that conclusions drawn from that study may fairly be combined with this into a single unified picture. Our initial galaxy selection is constructed using the Galaxy And Mass Assembly dataset. Catalogues produced by the GAMA team provide robust estimates for a number of galaxy properties, including aperture photometry, shape, concentration, stellar mass, dust corrected colour and, crucially, redshift. Target selections are sourced from the GAMA equatorial input catalogue Data Management Unit (DMU) tiling catalogue v46 \citep[TilingCatv46;][]{Baldry2010}. Matched aperture photometry across all $9$ passbands is provided in the aperture matched photometry DMU v06 \citep[ApMatchedCatv06;][]{Hill2011,Driver2016}. Robust automated measurements of \citet{Sersic1963,Sersic1968} light profiles used for the correction of matched Kron-like apertures to total flux are provided in the S\'{e}rsic photometry DMU SDSS S\'{e}rsic catalogue v09 \citep[SersicCatSDSSv09;][]{Kelvin2012}. Spectroscopic redshifts are taken from the GAMA spectroscopic DMU v27 catalogue \citep[SpecObjv27;][]{Liske2015} and subsequently corrected from heliocentric to local flow-corrected redshifts \citep{Tonry2000,Erdogdu2006} in the local flow correction DMU distances and reference frames v14 catalogue \citep[DistancesFramesv14;][]{Baldry2012}. 

Stellar mass and rest-frame dust-corrected colour estimates are provided in the stellar masses DMU v18 catalogue \citep[StellarMassesv18;][]{Taylor2011}. To derive these quantities, in brief, $u$ through $Y$ broadband photometric measurements are used to constrain stellar population synthesis (SPS) models for each galaxy. The models are based on \citet{Bruzual2003} stellar evolution models assuming a \citet{Chabrier2003} stellar initial mass function and a \citet{Calzetti2000} dust attenuation curve. The SPS fitting grid is accurate for galaxies in the range $z\leq0.65$. An earlier version of the distances and references frames catalogue (v13) has been used at this stage to determine distance, however, switching to v14 of the distances and references frames catalogue does not alter these outputs significantly. Outputs from this process include internal dust-corrected $u$ minus $r$ colour ($u^*-r^*$) and stellar mass estimates. These masses have been checked for consistency against mid-infrared stellar mass estimates based on high-quality data from the Wide-field Infrared Survey Explorer \citep[WISE,][]{Cluver2014,Kettlety2018}. An initial match of the stellar masses catalogue version 18 to the S\'{e}rsic photometry SDSS optical catalogue version 9 provides $198\,942$ galaxies. This comprises our master dataset. 

\begin{figure}
	\centering
	\includegraphics[width=\columnwidth]{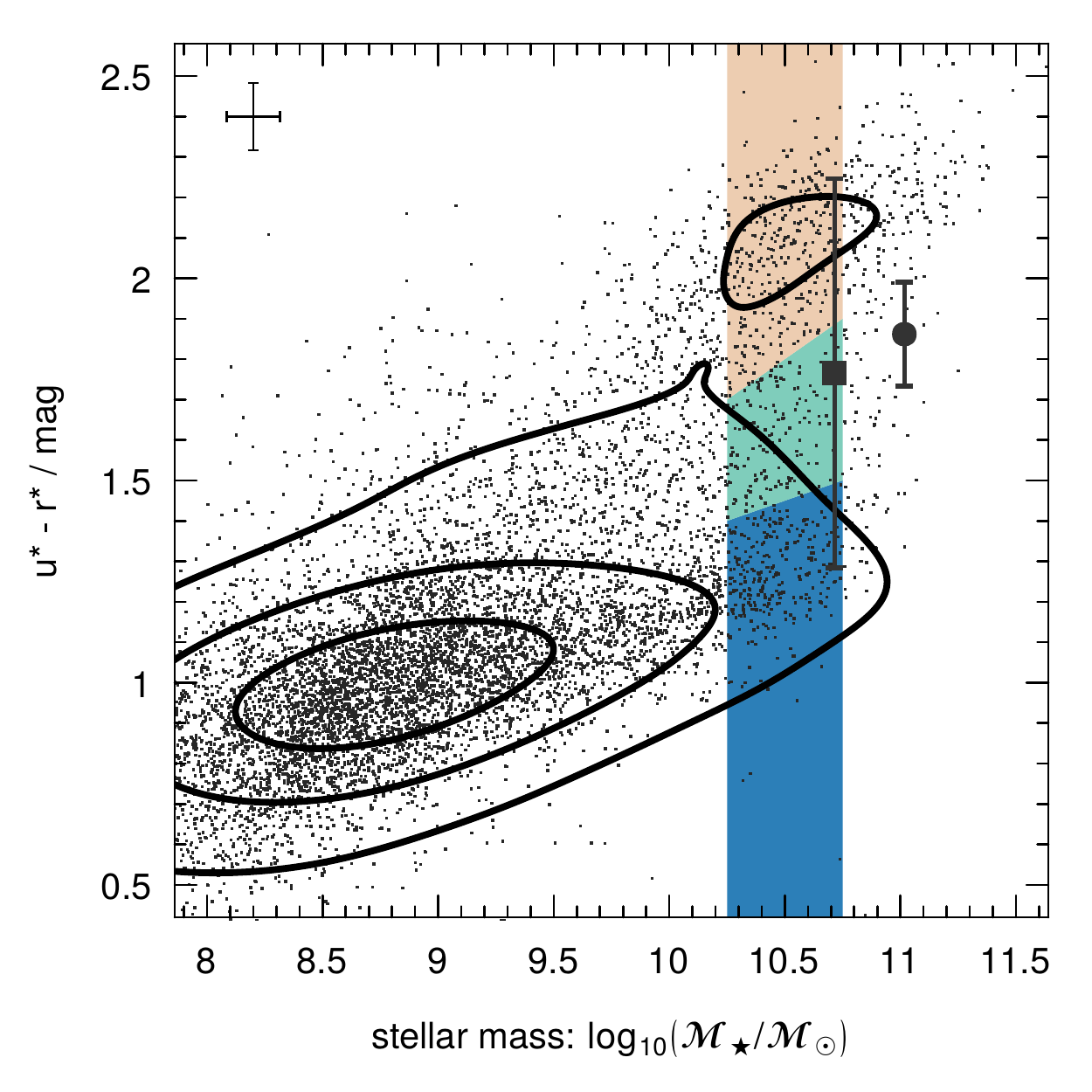}
    \caption{Rest-frame dust-corrected $u^*-r^*$ colour as a function of galaxy stellar mass. Data points show all galaxies in the master dataset which satisfy $z<0.06$, with solid black lines representing the 20/50/80 percentile contours. Our sample mass range of interest ($10.25<\log\mathcal{M}_{\star}/\mathcal{M}_{\odot}<10.75$) has been segregated into red, green and blue subsets, as indicated by colour. Median uncertainties for galaxies within this mass range are shown in the top-left corner. Local group values and confidence intervals for the Milky Way (square) and M31 (circle) have been added, for reference. See text for further details.}
    \label{fig:colmass}
\end{figure}

From our master sample, a low redshift\footnote{Here, redshift refers to \textit{either} heliocentric redshift \textit{or} Tonry local flow-corrected redshift following \cite{Baldry2012}, using boolean logic to select a maximally inclusive sample.} cut of $z<0.06$ is applied to guard against morphological observer bias at higher redshifts \citep{Bamford2009,Kelvin2014a,Hart2016}. These data are shown in the colour-mass plane in Figure \ref{fig:colmass} as black data points. Here we also show literature values and associated confidence intervals for the Milky Way (square) and M31 (circle)\footnote{Milky Way and M31 stellar mass confidence intervals are smaller than their respective data points.}, for reference \citep{Mutch2011,Bland-Hawthorn2016}. To account for the expected offset between intrinsic and observed colour at this stellar mass range, $u^*-r^*$ colour values for the Milky Way and M31 have been corrected downwards by $0.55$ mag \citep{Taylor2011}, placing both the Milky Way and M31 firmly within the green valley. In addition to a redshift cut, a further cut of $e<0.5$ (face-on\footnote{Ellipticity $e=1-b/a$, as defined via a S\'{e}rsic fit to the $r$-band KiDS image.}) is applied to the sample to minimise galaxy edge effects. The net result of these cuts is a reduced sample of $6\,272$ galaxies.

\subsection{Stellar Mass and Colour}
\label{sec:masscolour}

As in Paper I, we opt to explore the stellar mass range $10.25<\log\mathcal{M}_{\star}/\mathcal{M}_{\odot}<10.75$, shown in Figure \ref{fig:colmass} as the shaded coloured regions. This regime, within the stellar mass completeness limit for GAMA galaxies at this redshift range, is notable not only for residing close to the measured knee in the stellar mass function at $\log\mathcal{M}_{\star}/\mathcal{M}_{\odot}\sim10.6$ \citep[e.g.,][]{Baldry2008a,Peng2010b,Baldry2012,Peng2012,Kelvin2014b}, but is also the location of the high-mass tip of the blue cloud in the local Universe. Galaxies more massive than $\log\mathcal{M}_{\star}/\mathcal{M}_{\odot}=10.75$ increasingly tend to reside in the red sequence, implying that a redward transition for galaxies up to and inclusive of this mass range remains an important evolutionary mechanism. Furthermore, a colour transition at this relatively high stellar mass range necessarily implies a notable and visible impact on galaxy morphology. This makes this mass regime particularly interesting when studying any morphological and structural impacts on green valley transition. 

Two measures of stellar mass from the GAMA stellar masses catalogue are adopted: AUTO-defined (Kron-like) and S\'{e}rsic-defined. The former are stellar masses derived via stellar population synthesis modelling of $r$-defined AUTO (elliptical Kron) broadband photometry. The latter are a modification of the former using 2D S\'{e}rsic models to provide total flux corrections to the Kron-like aperture photometry. The effect of this correction is to correct for the known bias in recovered Kron-like fluxes as a function of Hubble type, and to provide a more accurate measure of total flux \citep{Graham2005a}. The flux ratio between AUTO and S\'{e}rsic apertures is given in StellarMassesv18 as \textit{fluxscale}. 

\begin{figure}
	\centering
	\includegraphics[width=\columnwidth]{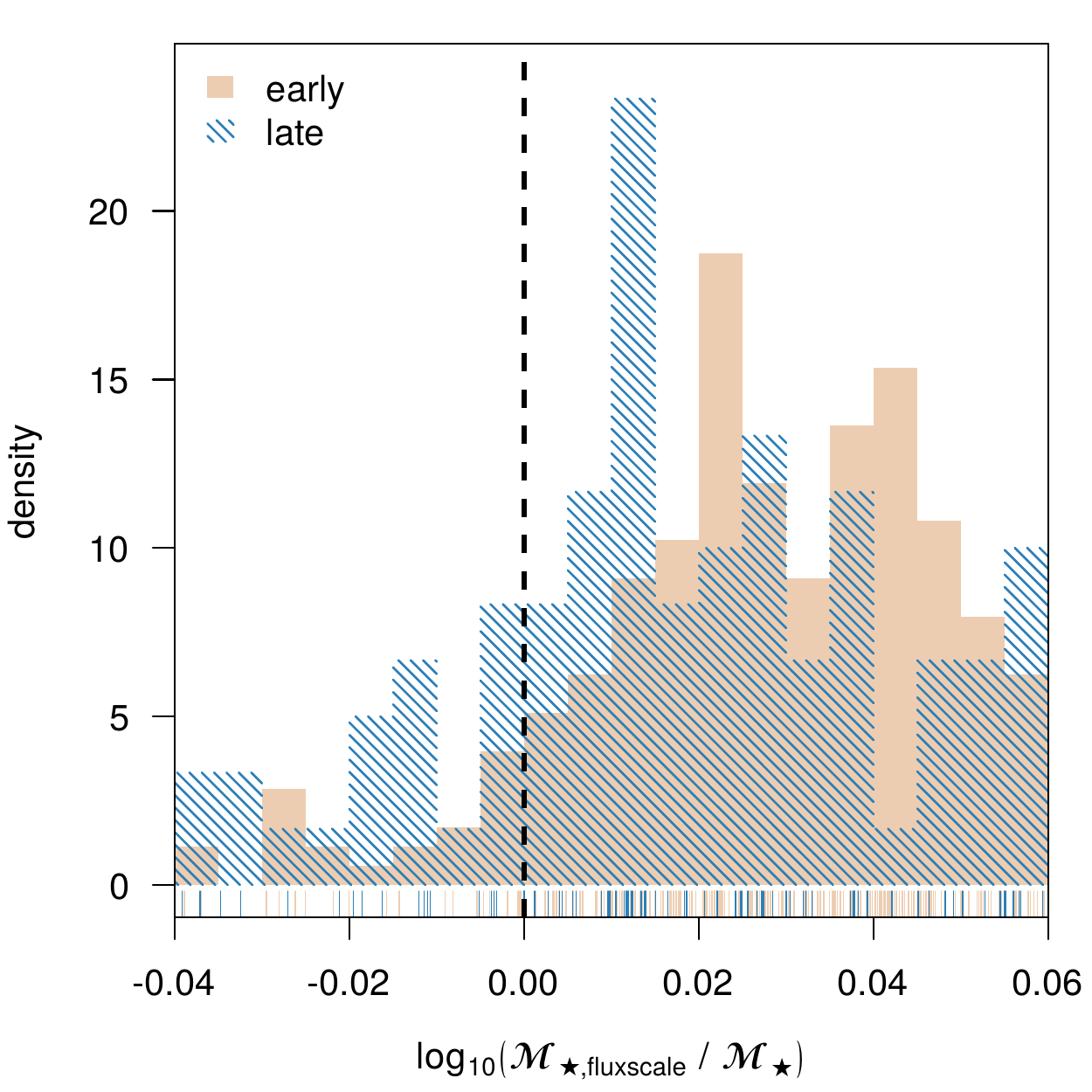}
    \caption{Stellar mass offset between S\'{e}rsic \textit{fluxscale} corrected stellar masses and AUTO Kron-defined stellar masses. Early- and late-type galaxies are shown as red and blue shaded regions, respectively.}
    \label{fig:fluxscale}
\end{figure}

To guard against extreme flux corrections, we limit \textit{fluxscale} to the range $0.9<\textit{fluxscale}<1.15$\footnote{Approximately $29\%$ of the total dataset exceed these limits.}. The result of these corrections to the resultant stellar masses can be seen in Figure \ref{fig:fluxscale}. This figure shows the stellar mass offset between S\'{e}rsic \textit{fluxscale} corrected stellar masses and AUTO-defined stellar masses for early- and late-type galaxies, respectively (see Section \ref{sec:morphologies} for further details on the determination of morphology). The resulting typical stellar mass correction corresponds to an average increase in stellar mass of the order $0.03$ dex. By Hubble type this corresponds to $0.02$ dex in late type galaxies, rising to $0.04$ dex in early type galaxies. To ensure complete coverage across our preferred mass regime where we properly sample both the red and blue populations, we select all galaxies for which \textit{either} definition of stellar mass falls in the range $10.25<\log\mathcal{M}_{\star}/\mathcal{M}_{\odot}<10.75$. Application of this cut reduces our sample of $6\,272$ to $505$ galaxies.

A match to the GAMA group finding DMU v09 \citep[catalogues G3CFoFGroupv09 and G3CGalv08;][]{Robotham2011b} provides group property measurements for these systems. Of our sample of $505$ galaxies, $265$ ($\sim52\%$) are ungrouped down to the GAMA flux limit of $r=19.8$. A further $173$ galaxies ($\sim34\%$) lie within low multiplicity groups consisting of $5$ members or fewer. Only the remaining $67$ galaxies ($\sim13\%$) fall within high multiplicity groups of more than $5$ members. This sample should therefore be considered as predominantly field-dominated.

Within this mass regime, we classify galaxies as ``red" if they are redder than a line connecting $u^*-r^*=1.7$ at $\log\mathcal{M}_{\star}/\mathcal{M}_{\odot}=10.25$, rising linearly to $u^*-r^*=1.9$ at $\log\mathcal{M}_{\star}/\mathcal{M}_{\odot}=10.75$. Similarly, ``blue" galaxies are defined as those bluer than a line connecting $u^*-r^*=1.4$ to $1.5$ over the same mass range. Galaxies between these two lines we term ``green". These boundaries have been selected to identify those regions in which the density of red sequence and blue cloud galaxies begins to drop off significantly. The boundaries we adopt here are somewhat redward of those applied to the higher redshift sample used in Paper I, indicating some small redshift dependence on tripartite colour division. Our selected boundaries subdivide our sample into $222$ red ($\sim44\%$), $94$ green ($\sim19\%$) and $189$ blue ($\sim37\%$) galaxies. Colour boundaries are shown in Figure \ref{fig:colmass} as the interface between the shaded red, green and blue regions. 

\subsection{Postage Stamps}
\label{sec:postage}

To facilitate visual morphological inspection, postage stamps for each galaxy are constructed using KiDS $g$- and $r$-band and VIKING $K$-band imaging data. Cutouts of size $50\times50$ kpc are produced for each galaxy in each filter. Images are convolved to the seeing of the worst band in $grK$ using a Gaussian filter with a full-width half-maximum $\Gamma$ defined as $\Gamma_{corr} = \sqrt[]{\Gamma_{max}^2 - \Gamma_{orig}^2}$. A synthetic band $x$ intermediate to $g$ and $r$ is defined as the linear arithmetic mean of the convolved $g$ and $r$ bands. Monochromatic $\arcsinh$-scaled postage stamps in $g$, $r$ and $K$ and a false 3-colour $\arcsinh$-scaled $rxg$ image are produced following the procedures outlined in \cite{Lupton2004}. Inverted versions of each image are also produced. The false 3-colour image should be understood to be the primary classification image, however, monochromatic information in all bands is produced to facilitate accurate morphological classification across a range of wavelengths. 

\begin{figure*}
	\centering
	\includegraphics[width=\textwidth]{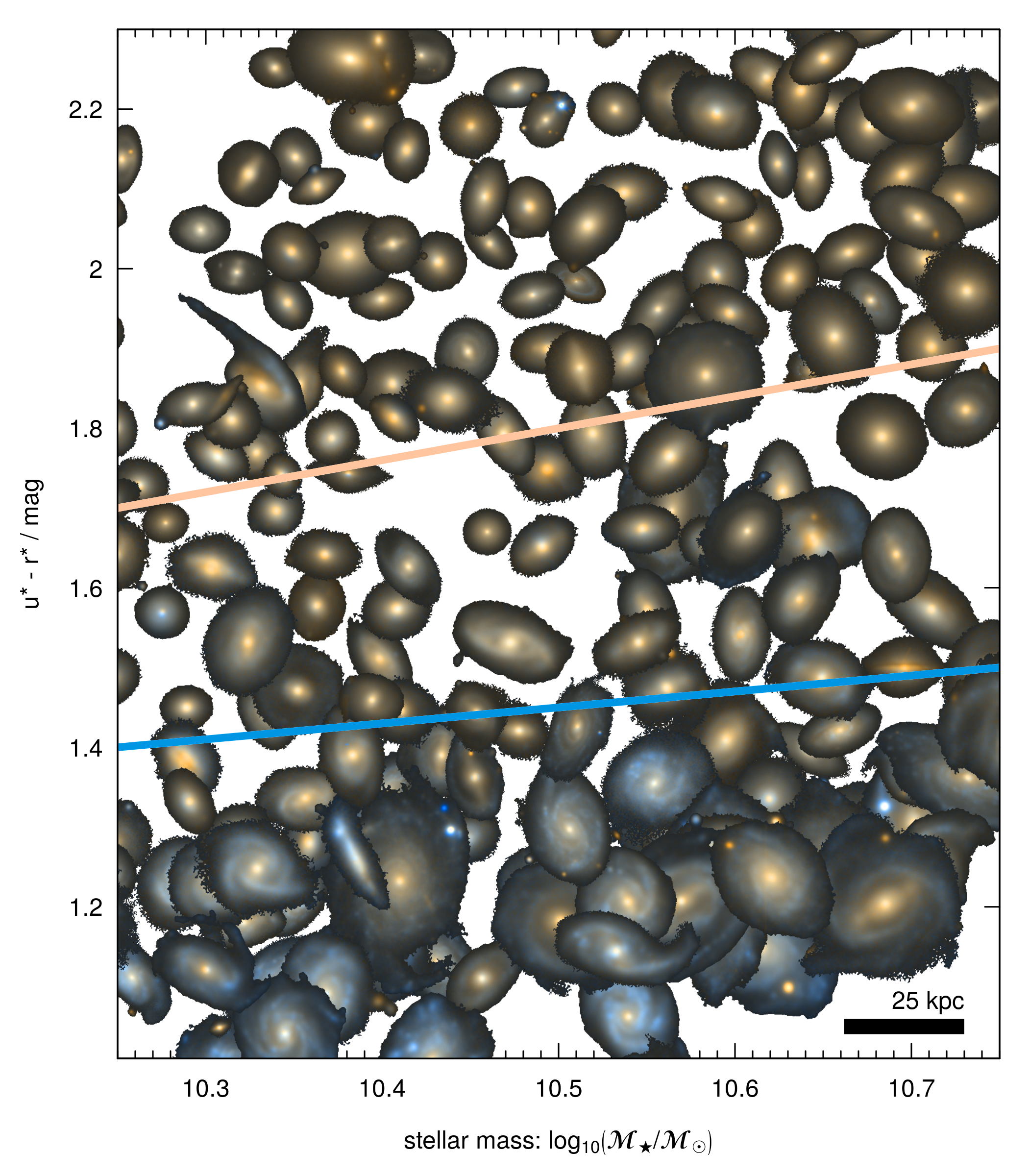}
    \caption{Postage stamps for $203$ representative galaxies in our final sample. These false 3-colour galaxy images are based on KiDS $g$ and $r$ band imaging. Galaxies are located at their reference position in the colour--stellar-mass plane (cf. Figure \ref{fig:colmass}). Solid red and blue lines indicate our chosen boundaries between the red, green and blue sub-populations. Images are scaled to physical units as represented by the inset legend.}
    \label{fig:postage}
\end{figure*}

Example 3-colour images using the above methodology for subsets of the red, green and blue populations are shown in Figure \ref{fig:postage} (cf. Figure \ref{fig:colmass}). Whilst the images shown in this figure have been trimmed to the outer isophote of each galaxy to aid in legibility, we confirm that the postage stamps ultimately used for classification do indeed extend to the full $50\times50$ kpc region. Galaxies are located at their reference position in the colour--stellar mass plane, with solid red and blue lines indicating the boundaries between the red, green and blue sub-populations. The boundaries of each image correspond to the same signal-to-noise cut. The increased density of objects along the blue cloud and red sequence can clearly be seen, with a much sparser population of objects across the green valley. As previously noted in the literature, the reduced density of objects across the green valley is seen as evidence for the relatively rapid transition of galaxies across this regime. We also note the apparent compactness of galaxies residing in the red sequence, relative to the relatively larger blue disks which populate the blue cloud.

\subsection{Morphologies}
\label{sec:morphologies}

Visual Hubble-type morphologies are assigned to each galaxy following the methodology outlined in \citet{Kelvin2014a}. The galaxy population is split into grouped Hubble types, namely, elliptical (E), (barred-)lenticular/early-type disk (S0-Sa), (barred-)intermediate-type disk (Sab-Scd) and late-type disk/irregular (Sd-Irr) galaxy classes based on visual inspection by LSK of $3$-colour KiDS postage stamp imaging. For the purposes of brevity, we shall refer to non-elliptical type galaxies (S0-Sa, Sab-Scd, Sd-Irr) as `disk-type' for the remainder of this paper. We opt to merge barred and unbarred classes into one over-arching class for both the S0-Sa and Sab-Scd populations. We find a good level of concordance between our morphological galaxy types recovered via this methodology and those of previous GAMA studies \citep{Kelvin2014a,Moffett2016a}, indicating the robustness of our classification approach. Of the $505$ galaxies in our sample set, $485$ return a match in the GAMA visual morphology v03 catalogue \citep{Kelvin2014a,Moffett2016a}. We note that $\sim11\%$ of galaxies previously classified as elliptical based on SDSS imaging have been reclassified as lenticular or early-type disk galaxies (S0-Sa) when re-analysed using deeper KiDS imaging. The misclassification of S0s as elliptical galaxies has been previously noted in the Carnegie Atlas of Galaxies, and similar results found more recently by \citet{Andrews2014} when comparing 2MASS, UKIDSS-LAS and VIKING NIR datasets, with shallower data struggling to accurately characterise galaxy structure at successively fainter magnitudes. The implications for this are important to bear in mind in an era of progressively deeper and ever improving astronomical imaging datasets.

On inspection of the postage stamps it was found that a number of galaxies suffer from potentially compromised photometry, requiring them to be removed from the sample. Typical criteria for exclusion are part of a galaxy incorrectly identified as the core, or a nearby bright neighbour/star significantly impacting the flux at the position of the galaxy centre. Each postage stamp is inspected by LSK, MNB and SP for evidence of contamination, and a cross check is made using the \textit{VIS\_CLASS} parameter in TilingCatv46 to identify targets with compromised photometry. Of our $505$ galaxies, $33$ suffer from compromised photometry and are consequently discarded. Removing these $33$ sources leaves us with our final dataset of $472$ galaxies: $213$ red ($\sim45\%$), $89$ green ($\sim19\%$) and $170$ blue ($\sim36\%$). The morphological properties of this sample are summarised in Table \ref{tab:sample}. Owing to the significant numbers of S0-Sa type galaxies within this sample, and similar to the distributions reported in Paper I, any trends reported for disk-type galaxies should be understood to be heavily influenced by trends in the S0-Sa class. This sample will be used throughout the remainder of the paper.

\begin{table}
	\centering
	\begin{tabular}{l|r@{\hspace{0cm}}l@{\hspace{0cm}}r@{\hspace{0cm}}l@{\hspace{0.2cm}}r@{\hspace{0cm}}l@{\hspace{0cm}}|r@{\hspace{0cm}}l@{\hspace{0cm}}}
		Hubble Type & Red && Green && Blue && Total &\\
		\hline
		E        	& $87$  && $27$ && $10$ && $124$ &\\
        S0-Sa 		& $121$ & \multirow{3}{*}{$\begin{rcases*}\\\\\end{rcases*}126$} & $51$ & \multirow{3}{*}{$\begin{rcases*}\\\\\end{rcases*}62$} & $56$ & \multirow{3}{*}{$\begin{rcases*}\\\\\end{rcases*}160$} & $228$ & \multirow{3}{*}{$\begin{rcases*}\\\\\end{rcases*}348$}\\
        Sab-Scd  	& $3$   && $11$ && $103$ && $117$ &\\
        Sd-Irr   	& $2$   && $0$ && $1$ && $3$ &\\
        \hline
        Total 		& $213$ && $89$ && $170$ && $472$ &\\
	\end{tabular}
    \caption{Breakdown by Hubble type and colour for all $472$ galaxies in our final sample. Braces show total numbers for disk-type galaxies (S0-Sa, Sab-Scd and Sd-Irr).}
	\label{tab:sample}
\end{table}

\section{Green Valley Census}
\label{sec:census}

To facilitate a more detailed morphological classification, we make use of the Zooniverse\footnote{www.zooniverse.org} \textit{Build A Project} classification tool. This is an online web resource which enables teams of people to classify images based on a particular question set. Galaxy classification via consensus has become an increasingly popular tool in recent years \citep[for example, see][]{Lintott2008,Baillard2011,Willett2013,Hart2016}, providing a simple and effective means by which large samples of galaxies may be processed. On the Zooniverse platform we constructed the \textit{Green Valley Census} with the aim to quantify galaxy morphological indicators. We construct a simple decision tree as follows:

\begin{enumerate}  
\item Does the galaxy contain a bar, and if so, how strongly barred is it?
\begin{itemize}
\item No bar
\item Weakly barred
\item Strongly barred
\end{itemize}
\item Do you see any of these features?
\begin{itemize}
\item Ring / partial ring
\item Plateau / lens
\item Tidal tails / streams
\item A shell / shells
\item Interaction / merger
\end{itemize}
\end{enumerate}

Classifiers may select only one option from question (i) and multiple options from question (ii) as appropriate. Most structures require no explanation as to their nature. Guidance was given with regard to the `Plateau/lens' category \citep[see][also Section \ref{sec:introduction}]{Kormendy1979}. These types of structures are believed to be dynamically linked with (dissolving) bars or tied to evolved rings. Lenses are distinct features having a shallow surface brightness gradient interior to a sharp edge. Images are uploaded to the Green Valley Census in $g$, $r$ and $K$ monochrome bands as well as a 3-colour $rxg$ image (see section \ref{sec:postage}) for all $472$ galaxies\footnote{Historically, galaxy classification was based on blue light imaging, rather than the more comprehensive multi-wavelength imaging in use here. The classifications we report may therefore differ somewhat from those available in earlier literature.}.

A total of $8$ classifiers\footnote{Namely, in alphabetical order: AJM, LJMD, LSK, MNB, PAJ, RDP, SMP and SP.} classified the entire sample. In the rare event that a classifier provided multiple classifications for the same object, the most recent classification is taken and the older discarded. During the data reduction process it became apparent that the definition of `strongly barred' and `weakly barred' is somewhat subjective. To that end, we reduce question (i) in our subsequent analysis by combining these two options into a simple `barred' category. Overall, a good level of agreement is found between classifiers, with the incidence of extreme outliers (a classifier being the only one to classify a galaxy a certain way, or the only one \textit{not} to classify a certain way) being rare, typically occurring in $\lesssim$$1\%$ of cases (see Figure \ref{fig:userstats}). This indicates a good level of concordance between classifiers for most systems, underlining the strength of our multi-classifier approach towards galaxy feature identification. Further discussion of classifier concordance and classification outliers may be found in Appendix \ref{outliers}. The complete Green Valley Census catalogue showing anonymised user votes for each structural indicator is available online via the VizieR database of astronomical catalogues at the Centre de Donn\'{e}es astronomiques de Strasbourg (CDS) website\footnote{Available at CDS via anonymous FTP to cdsarc.u-strasbg.fr (130.79.128.5)
or via http://cdsarc.u-strasbg.fr/viz-bin/qcat?J/MNRAS}.

\section{Results}
\label{sec:results}

\begin{figure*}
	\centering
	\includegraphics[width=\textwidth]{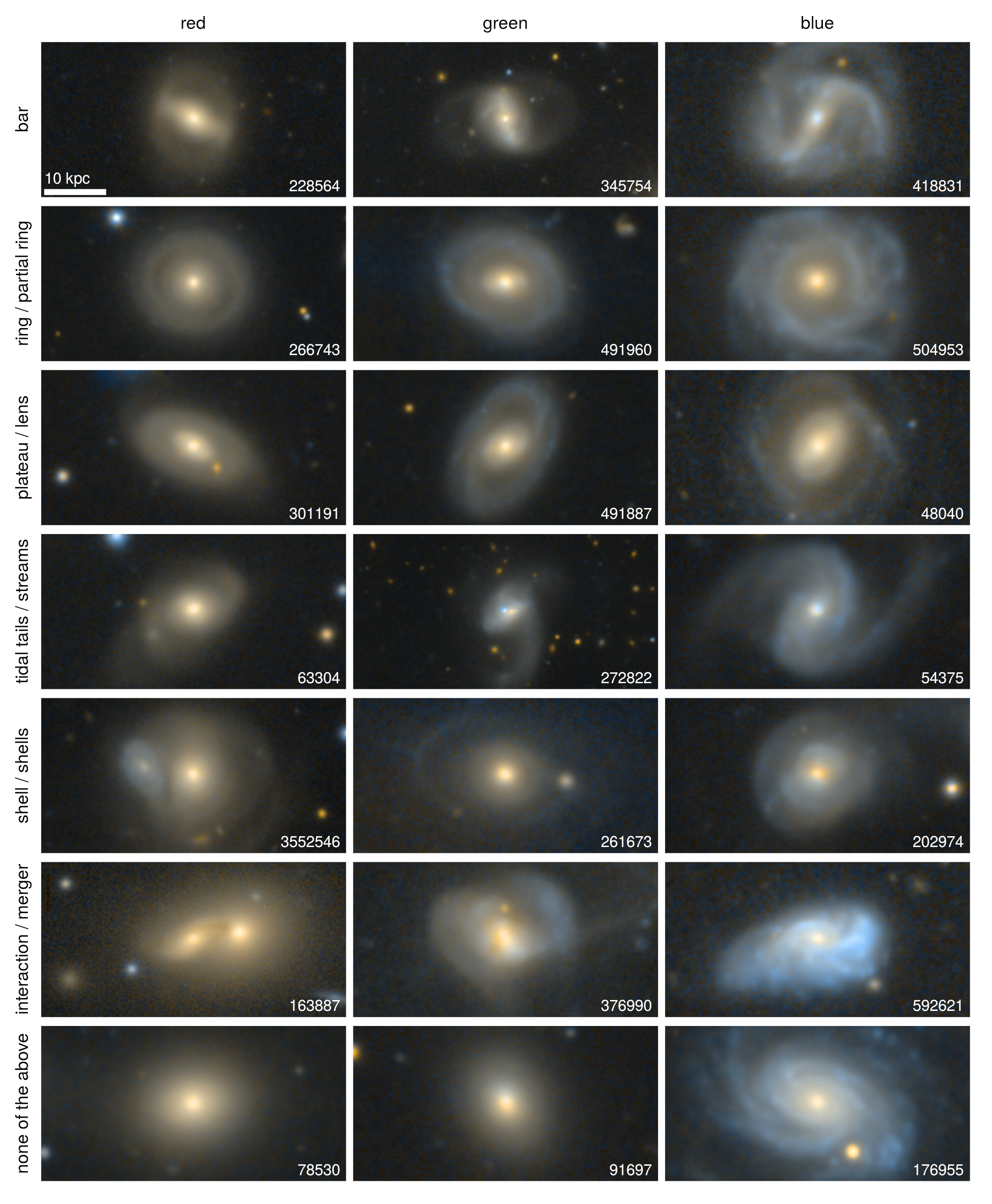}
    \caption{Examples for each of the six structural components classified (top six rows), separated into red, green and blue, as indicated. The bottom row shows example galaxies in which none of the six structural markers were identified. Postage stamps are constructed using KiDS $g$ and $r$ band imaging, $\arcsinh$ scaled. Each galaxy is shown at the same physical scale as represented by the scale bar in the top left panel. The GAMA CATAID of the pictured galaxy is given in the lower right corner of each panel.}
    \label{fig:examples}
\end{figure*}

Results from our Green Valley Census are reduced into a final catalogue as follows. Based on the number of votes given by classifiers, each galaxy is assigned a score from $0$ to $8$ for each of the six structural indicators shown in Section \ref{sec:census}, namely: bar, ring/partial ring, plateau/lens, tidal tails/streams, shell/shells and interaction/merger. Those galaxies with $4$ or more votes in any given category are assigned to that category, i.e., at least half of the classifiers must have classified the galaxy in that way. Our elliptical population shows no significant trends for many of these structural markers, as might be expected, and therefore we limit our further analyses to disk-type galaxies alone (S0-Sa, Sab-Scd and Sd-Irr). This disk-type population is split by colour into red, green and blue sub-groups as discussed in Section \ref{sec:masscolour}. A vote fraction is determined for each sub group with fractional errors assigned following the beta distribution quantile technique \citep{Cameron2011}. Errors derived in such a fashion are consistent with other techniques (e.g., normal approximation) when applied to large datasets, yet have been shown by \citet{Cameron2011} to outperform such traditional methodologies for small data samples. Example galaxies meeting the above classification criteria for each of the six structural components are shown in Figure \ref{fig:examples}, with the final row providing examples for which none of the above structures were noted. 

\subsection{Variation in Structure as a Function of Galaxy Colour}
\label{sec:mainresults}

\begin{figure*}
	\centering
	\includegraphics[width=\textwidth]{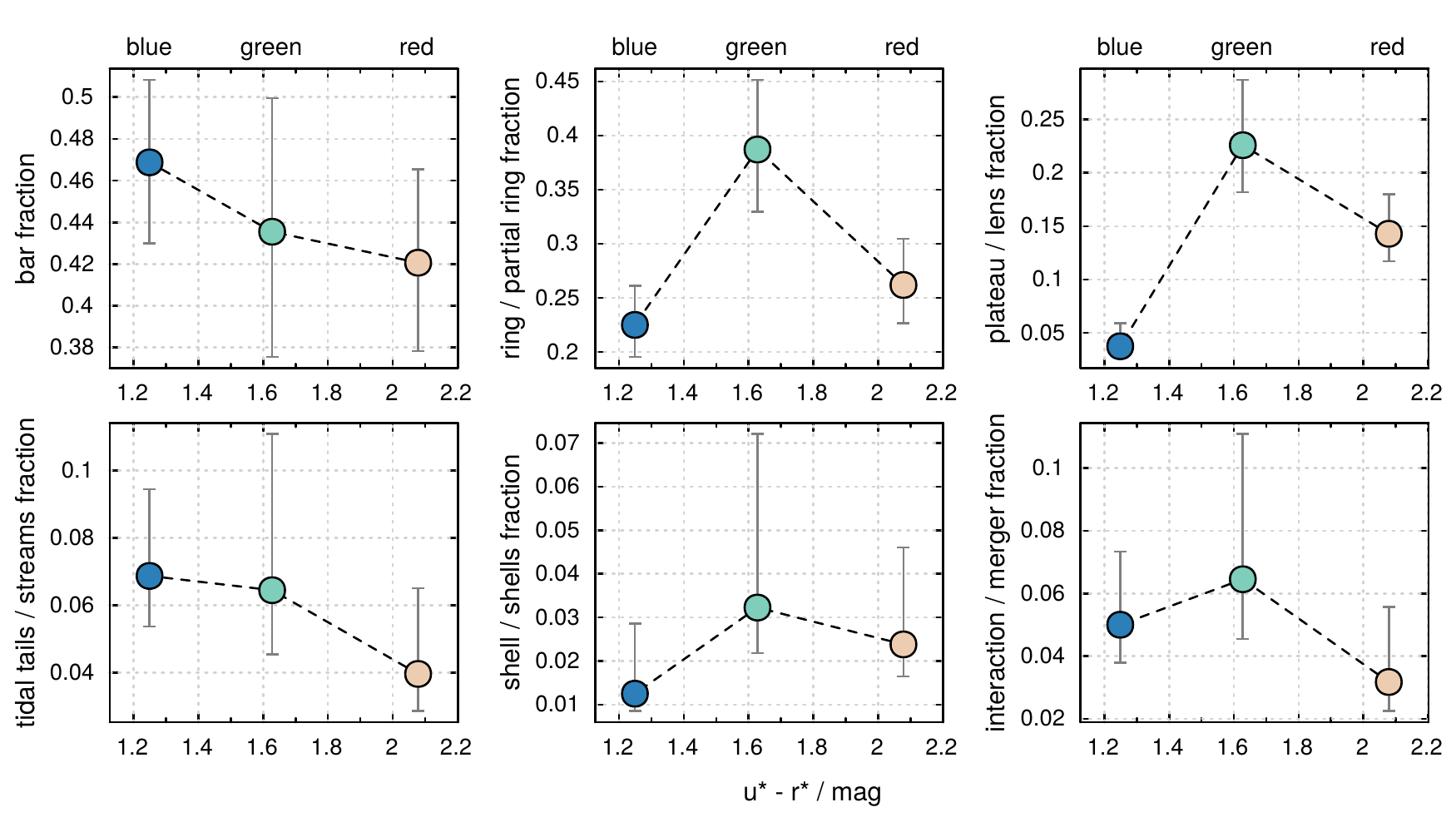}
    \caption{Number fractions for various structural indicators as a function of $u^*-r^*$ colour for disk-type galaxies. Error bars represent $1\sigma$ confidence intervals. These data are shown in tabulated form in Table \ref{tab:non-elliptical}.}
    \label{fig:structures}
\end{figure*}

Our primary classification results are shown in Figure \ref{fig:structures}. Each panel shows the number fraction of the disk-type population as a function of $u^*-r^*$ colour for a given structural indicator. Fractional $1\sigma$ errors are estimated as previously discussed. The position of the data points with respect to the $x$-axis corresponds to the median colour for that particular colour group. Many indicators are consistent with a flat or nearly-flat progression across the green valley within their errors, however, a number of results such as the disk-type galaxy ring and lens fractions warrant further investigation.

The results presented in Figure \ref{fig:structures} are shown in tabulated form in Table \ref{tab:non-elliptical}. Each row represents a sub-division by colour, as indicated, whilst $\mathrm{N_{tot}}$ shows the total number of galaxies of a given colour as in Table \ref{tab:sample}. The remaining columns show the percentage number fraction for a given class, with the final column a combination of ring and lens types into a single class. Values within parentheses show the $\sigma$ significance of the green valley number fraction relative to the weighted average of the red/blue population. 

\begin{table*}
    \centering
    \begin{tabular}{l|l|lllllll}
        \multirow{2}[5]{*}{colour} &
        \multirow{2}[5]{*}{$\mathrm{N}_\mathrm{tot}$} &
        \multicolumn{7}{c}{number fraction (\%)}\\ \cmidrule(lr){3-9} && bar & ring & lens & tidal & shell & merger & ring/lens\\
        \hline
        red & 126 & $42.06_{-4.24}^{+4.49}$  & $26.19_{-3.52}^{+4.27}$  & $14.29_{-2.57}^{+3.69}$  & $3.97_{-1.10}^{+2.54}$  & $2.38_{-0.73}^{+2.23}$  & $3.17_{-0.93}^{+2.40}$  & $37.30_{-4.08}^{+4.48}$ \\[5pt]
        green & 62 & $43.55_{-6.00}^{+6.40}$ ($0.2$) & $38.71_{-5.74}^{+6.45}$ ($2.3$) & $22.58_{-4.41}^{+6.13}$ ($2.9$) & $6.45_{-1.90}^{+4.63}$ ($0.3$) & $3.23_{-1.04}^{+3.98}$ ($0.9$) & $6.45_{-1.90}^{+4.63}$ ($0.9$) & $53.23_{-6.34}^{+6.14}$ ($3.3$)\\[5pt]
        blue & 160 & $46.88_{-3.88}^{+3.96}$  & $22.50_{-2.95}^{+3.63}$  & $3.75_{-0.99}^{+2.14}$  & $6.88_{-1.50}^{+2.57}$  & $1.25_{-0.40}^{+1.61}$  & $5.00_{-1.21}^{+2.33}$  & $25.00_{-3.10}^{+3.72}$ \\
    \end{tabular}
    \caption{Number fractions for all structural indicators found within disk-type galaxies. Each row represents a sub-division by colour. $\mathrm{N_{tot}}$ shows the total number of galaxies of a given colour. The remaining columns show the percentage number fraction for a given class, with the final column a combination of ring and lens types. Fractional $1\sigma$ errors are estimated via the beta distribution quantile technique. Values within parentheses show the $\sigma$ significance of the green valley number fraction offset relative to the weighted average of the red/blue population.}
    \label{tab:non-elliptical}
\end{table*}

Of significant interest are ringed and lens type number fractions for disk-type galaxies. The ring fraction is $\sim23\%$ in the blue, rising to $\sim39\%$ in the green before falling back to $\sim26\%$ in the red. Similarly, the lens fraction is $\sim4\%$ in the blue, rising to $\sim23\%$ in the green before falling back to $\sim14\%$ in the red. The significance in the surplus of rings and lenses across disk-type green valley galaxies is $2.3\sigma$ and $2.9\sigma$, respectively.

The number fractions for tidal streams, shells and interacting/merging galaxies are all relatively low, typically at $\lesssim7\%$ for all colours. The trends we recover are consistent with a flat or almost flat relation with colour, implying little to no structural impact across the green valley. In the case of streams and shells, this may be due to difficulties in extracting and preserving low surface brightness (LSB) flux. Streams and shells become increasingly evident at $\mu_r>30\,\mathrm{mag}\,\mathrm{arcsec}^{-2}$, however, this regime is notoriously difficult to fully exploit, often requiring specialised LSB flux detection algorithms \citep[e.g.,][]{Williams2016}. LSB flux may easily be contaminated or destroyed, e.g., by scattered light from nearby bright sources or sky over-subtraction. In the case of interacting galaxies we note that our sample is small, relatively low-redshift, and we remind the reader that this sample should be considered to consist of predominantly field galaxies. To that end, our recovered low interaction fraction is not entirely unexpected and remains consistent with results from previous studies \citep[e.g.,][]{Knapen2009,Robotham2011b,DePropris2014}.

Finally, our recovered bar fraction for disk-type galaxies is also consistent with being flat with colour, with a value of $\sim44\%$. This is in good agreement with the literature for low redshift galaxies of this mass \citep[e.g.,][]{Sheth2008}. No significant offset is noted for those galaxies transitioning the green valley, with a green valley surplus significance of $0.2\sigma$.

\subsection{Rings and Lenses}
\label{ringlens}

\begin{figure}
	\centering
	\includegraphics[width=\columnwidth]{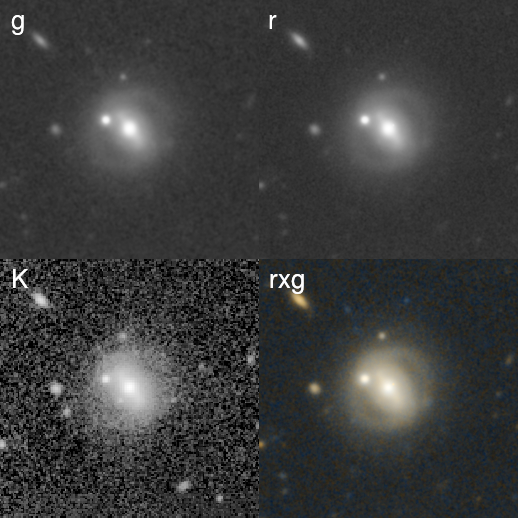}
    \caption{G422286 as observed in KiDS $g$ and $r$ bands (top left and right, respectively), the VIKING $K$ band (bottom left), and a 3-colour $rxg$ image (bottom right). Postage stamps are $\arcsinh$ scaled, each spanning $50\times50$ kpc in size. Note the distinct ring-like feature in the optical is washed out in the NIR, appearing instead as an outer lens.}
    \label{fig:ringlens}
\end{figure}

The recovered number fractions for ringed and lens-type galaxies show similar trends as a function of colour, namely, a surplus across the green valley relative to both the blue cloud and the red sequence. Visual inspection of many of these systems indicates that an optical ring tends to appear lens-like when observed in the near infrared. Figure \ref{fig:ringlens} shows galaxy G422286 as observed in the optical KiDS $g$ and $r$ bands (top left and top right respectively) and the NIR VIKING $K$ band (bottom left). An outer ring is visible in the optical, connecting at or close to the tip of the bar. However, in the NIR the presence of the ring is less clear, instead bearing a close resemblance to a diffuse outer morphological lens. G422286 received both exclusive ring and lens votes, with classifier confusion undoubtedly contributing to the splitting of the vote across these two linked structures \citep[see also][]{Kormendy1979}. Whilst this scenario is rare, the potential for confusion between ring and lens classifications should be noted.

\begin{figure}
	\centering
	\includegraphics[width=\columnwidth]{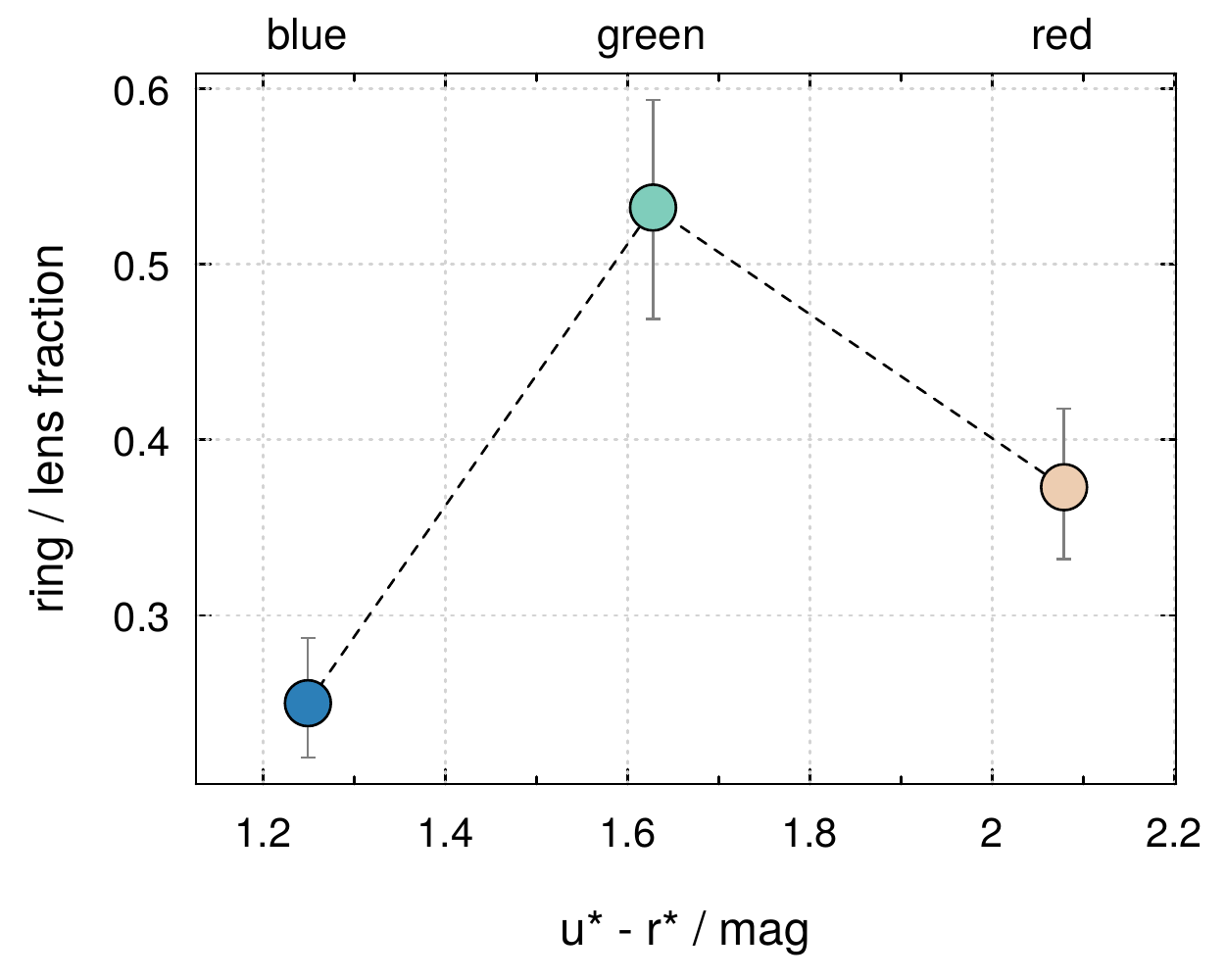}
    \caption{Number fraction for recovered ring/lens type structures as a function of $u^*-r^*$ colour for disk-type galaxies. Error bars represent $1\sigma$ confidence intervals.}
    \label{fig:circle}
\end{figure}

To mitigate the effects of structural mis-classification (owing to similarities between ring and lens-type galaxies) dependent upon observed wavelength, we opt to merge ring and lens votes into a single over-arching category. Figure \ref{fig:circle} shows the combined ring/lens number fraction as a function of $u^*-r^*$ colour for our entire sample. These results are displayed in tabulated form in the final column of Table \ref{tab:non-elliptical} for our disk-type galaxy population. Note that the combined ring/lens number fractions will, by design, be less than or equal to the summation of the separate ring and lens number fractions, owing to the fact that classifiers classified some small proportion of galaxies as exhibiting both a ring and a lens. A distinct surplus in the recovered ring/lens fraction is found in the green valley relative to both the blue cloud and the red sequence. We find $\sim25\%$ ring/lens type galaxies in the blue, rising to $\sim53\%$ in the green before falling back to $\sim37\%$ in the red. The significance of the surplus of ring/lens type structures transitioning across the green valley is $3.3\sigma$.

\subsection{Influence of a Bar}
\label{sec:bar}

\begin{figure}
	\centering
	\includegraphics[width=\columnwidth]{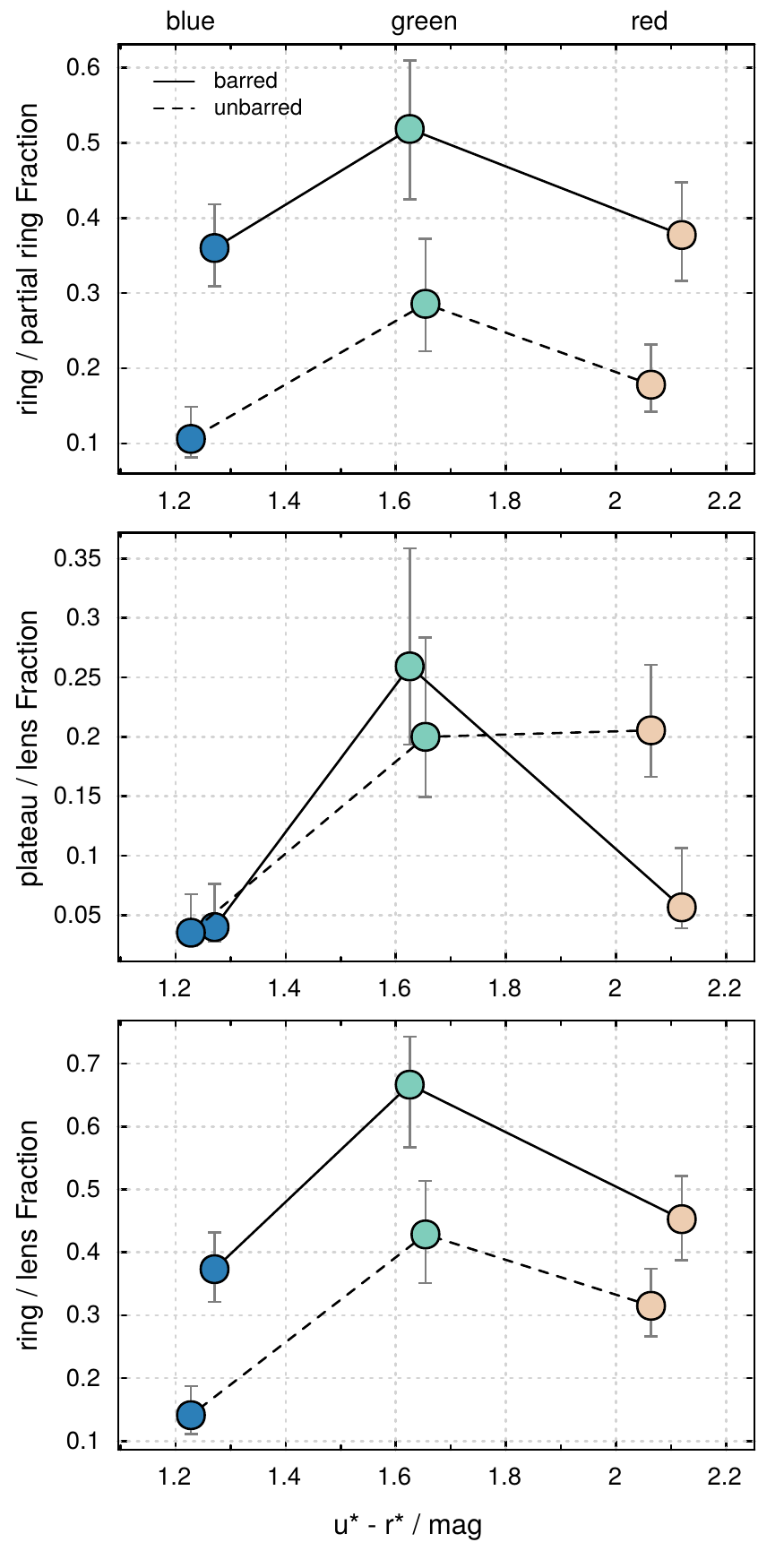}
    \caption{Number fraction for recovered rings (top), morphological lenses (middle) and combined ring/lens type structures (bottom) as a function of $u^*-r^*$ colour for disk-type galaxies. The galaxy population has been split into barred (solid line) and unbarred (dashed line) subsamples. Error bars represent $1\sigma$ confidence intervals.}
    \label{fig:bar}
\end{figure}

We have shown that disk-type galaxy bar fractions show no significant surplus across the green valley, however, a bar may still influence other structural features. We explore what influence the presence or otherwise of a bar has on the recovered ring and/or lens number fractions. Figure \ref{fig:bar} shows the ring (top), lens (middle) and combined ring/lens (bottom) number fractions for disk-type galaxies as a function of $u^*-r^*$ colour. The solid line shows the trend for barred galaxies, whilst the dashed line shows the trend for unbarred galaxies. These results are shown in tabulated form in Tables \ref{tab:non-elliptical+barred} and \ref{tab:non-elliptical+unbarred} for disk-type barred and unbarred galaxies, respectively. 

%NON-ELLIPTICAL\+BARRED+UNBARRED
\begin{table}
    \centering
    \begin{tabular}{l|l|lll}
        \multirow{2}[5]{*}{colour} &
        \multirow{2}[5]{*}{$\mathrm{N}_\mathrm{tot}$} &
        \multicolumn{3}{c}{number fraction (\%)}\\ \cmidrule(lr){3-5} && ring & lens & ring/lens\\
        \hline
        red & 53 & $37.74_{-6.10}^{+6.99}$  & $5.66_{-1.76}^{+5.00}$  & $45.28_{-6.54}^{+6.88}$ \\[5pt]
        green & 27 & $51.85_{-9.41}^{+9.15}$ ($1.5$) & $25.93_{-6.58}^{+9.92}$ ($3.0$) & $66.67_{-9.96}^{+7.65}$ ($2.4$)\\[5pt]
        blue & 75 & $36.00_{-5.11}^{+5.84}$  & $4.00_{-1.24}^{+3.64}$  & $37.33_{-5.19}^{+5.85}$ \\
    \end{tabular}
    \caption{Number fractions for rings, morphological lenses and a combined ring/lens class for disk-type barred galaxies. Each row represents a sub-division by colour. $\mathrm{N_{tot}}$ shows the total number of galaxies of a given colour. The remaining columns show the percentage number fraction for a given class. Errors are estimated via the beta distribution quantile technique. Values within parentheses show the $\sigma$ significance of the green valley number fraction relative to the weighted average of the red/blue population.}
    \label{tab:non-elliptical+barred}
    \begin{tabular}{l|l|lll}
        \multirow{2}[5]{*}{colour} &
        \multirow{2}[5]{*}{$\mathrm{N}_\mathrm{tot}$} &
        \multicolumn{3}{c}{number fraction (\%)}\\ \cmidrule(lr){3-5} && ring & lens & ring/lens\\
        \hline
        red & 73 & $17.81_{-3.61}^{+5.33}$  & $20.55_{-3.92}^{+5.50}$  & $31.51_{-4.88}^{+5.87}$ \\[5pt]
        green & 35 & $28.57_{-6.32}^{+8.65}$ ($2.1$) & $20.00_{-5.08}^{+8.34}$ ($1.5$) & $42.86_{-7.75}^{+8.53}$ ($2.4$)\\[5pt]
        blue & 85 & $10.59_{-2.47}^{+4.29}$  & $3.53_{-1.09}^{+3.24}$  & $14.12_{-2.97}^{+4.62}$ \\
    \end{tabular}
    \caption{As Table \ref{tab:non-elliptical+barred}, but for disk-type unbarred galaxies.}
    \label{tab:non-elliptical+unbarred}
\end{table}

A surplus of rings, lenses and ring/lens type galaxies across the green valley is similarly reproduced in both barred and unbarred systems. Interestingly, in the case of rings and combined rings/lenses, these structures are more numerous by $\sim20-30$ percentage points in barred systems than in their unbarred counterparts for all colour groups. For disk-type unbarred galaxies we find ring/lens fractions of $\sim14\%,\sim43\%,\sim32\%$ in blue, green, red, respectively, with a green valley surplus significance of $2.4\sigma$. For disk-type barred galaxies we find ring/lens fractions of $\sim37\%,\sim67\%,\sim45\%$ in blue, green, red, respectively, with a green valley surplus significance also of $2.4\sigma$.

The variation across the green valley for lenses in disk-type galaxies shows a different trend to that for rings. Barred disk-type galaxies exhibit a significant $3.0\sigma$ surplus of lenses relative to their blue/red counterparts. A weaker trend is found for the equivalent unbarred systems, with a reduced green valley surplus significance of $1.5\sigma$. This weakness is due to the similar fractions of lens galaxies for green valley and red sequence non-barred galaxies.

\section{Discussion}
\label{sec:discussion}

Exploring the redshift and stellar mass range $z<0.06$ and $10.25<\log\mathcal{M}_{\star}/\mathcal{M}_{\odot}<10.75$, respectively, we find a distinct surplus of ring and morphological lens type structures with morphologies that include a disk component as they transition across the green valley. As shown in Figure \ref{fig:structures}, this equates to a green valley ring surplus of $\sim10-15$ percentage points and a green valley lens surplus of $\sim10-15$ percentage points relative to either the blue cloud or the red sequence. Indeed, as the green valley tends to be dominated by lenticular/early-type disk galaxy classes (Table \ref{tab:sample}), then it appears that these trends also somewhat correlate with morphology, in agreement with previous studies \citep[e.g.,][]{Comeron2014}. The significance of this green valley surplus is $2.3\sigma$ in the case of rings, and $2.9\sigma$ in the case of lenses, with a combined ring plus lens green valley surplus significance of $3.3\sigma$. It is unlikely that catastrophic events (e.g. major mergers) that would act to destroy such features are the primary mode of evolution across the green valley in the majority of cases.

As shown in Figure \ref{fig:colmass}, both the Milky Way and M31 appear to reside within the green valley. Previous studies have shown that the Milky Way exhibits an X-shaped bulge \citep{Saito2011}. Following \citet{Salo2017} and \citet{Laurikainen2017}, when viewed face on this X-shaped bulge likely resembles a lens-like barlens structure. M31 has been shown to possess both an inner and outer ring, believed to have formed via a direct head-on collision with M32 \citep{Block2006}. Furthermore, \citet{Pagani1999} report that the star-formation rate of M31 is uniquely low, even in the ring, relative to equivalent local group spiral galaxies. It is therefore interesting to note that both of these local group green valley galaxies, with strikingly low star-formation rates in the case of M31, possess both of the structural features we find in relative excess for our own sample of green valley galaxies. 

The cause of such structural distinctiveness for galaxies within the green valley remains unclear. \citet{Mendez2011} found the morphologies of green valley galaxies to be intermediate to those of their red and blue counterparts, counter to what we observe here. If the dominant mode of travel is from the blue cloud to red sequence, one possible scenario is that these structures must be formed and then subsequently destroyed as they enter and then exit the green valley. As previously highlighted, \citet{Kormendy1979} postulated that a dissolving bar may lead to the formation of a morphological lens. Using a series of N-body simulations, \citet{Athanassoula2013} found that structures such as bars form more easily in gas-poor systems, which may explain the sudden surplus of bar-associated features such as rings and lenses in galaxies beginning to exhaust their available gas supply (as posited in Paper I). However, whilst we do find a surplus of bar-related structures such as rings and lenses in the green valley, we do not find an equivalent surplus of bars themselves, as might be expected if the aforementioned scenario applies here.

An alternative hypothesis is that these ringed and lens-like structures may always have been present in the host disk galaxy from a prior evolutionary mechanism, yet have, until now, remained undetectable beneath the emission of the relatively brighter disk. In this scenario it is the process of green valley transition itself which temporarily exposes such structures as the disk light profile flattens out due to a reduction in or cessation of star-formation initially in its central regions before ultimate fading of the ring/lens acts to re-mask these features. If this assumption is correct, then the ring and lens number fractions for galaxies in the green valley are closer to what one might consider to be the `true' number fractions, whilst the number fractions in either the blue cloud or the red sequence represent `masked' or otherwise attenuated number fractions. 

Both proposed mechanisms above agree well with the so-called `inside-out' death of star formation hypothesis \citep[e.g.,][]{Perez2013}. This contends that star-formation is suppressed initially in the core regions of a galaxy, and the suppression slowly propagates outwards as the galaxy exhausts its gas reservoir. In the early stages of this process, the centre of a disk will fade relative to its outskirts. This process may be linked to the dominance of the central bulge, as evidenced by the preponderance of S0-Sa type galaxies within our green valley sample, which perhaps indicates the presence of a strong bulge as a necessary precondition before entry into the green valley \citep[Paper I, see also][]{Fang2013}. Possible central quenching mechanisms include the creation of a bar which then acts to sweep out the central region \citep[cf. `Star Formation Desert'][]{James2015,James2016,James2018}, morphological quenching of the gaseous disk \citep{Martig2009} or quenching due to the presence of an AGN, potentially formed via the funnelling of gas along the bar into the central nexus. A sufficiently faded inner disk may come to resemble a ring or lens-type structure, supporting our former green valley surplus scenario, or may act to reveal these structures if they were already in existence. Furthermore, we can rule out catastrophic mechanisms as the likely cause of these noted trends, as any catastrophic quenching mechanism (e.g., a major merger) would likely destroy or disturb any ring or lens-like structure.

The bar also plays a crucial role in modifying our recovered trends. Figure \ref{fig:bar} shows that the presence of a bar correlates strongly with a surplus of observed rings regardless of colour. A link between bars and colour has previously been noted in \citet{Kruk2018}, who find that the disks of unbarred galaxies are significantly bluer than the disks of barred galaxies. This builds upon work by \citet{Masters2010} who indicate a potential link between the bar and the cessation of star formation in spiral systems. The area swept out by a bar, the Star Formation Desert, likely aids in the definition and recognition of an inner ring in these ringed systems. The presence of a bar also has a strong green valley effect on the detection of morphological lenses, with a green valley surplus significance of $3.0\sigma$. If not a statistical anomaly, it is curious that the effect of the bar here impacts the green valley alone, whereas the impact of the bar on ring-like structures impacts red, green and blue equally. A lens surplus across the green valley suggests a dynamical link between the exhaustion of a galaxy's gas reservoir, the formation or early phases of dissolution of a bar, and the formation of a morphological lens. It is well-known that bars in late-type galaxies are fundamentally different from bars in earlier types \citep{Elmegreen1985}. If galaxies generally evolve from the blue cloud into the red sequence, the nature of their bars must also change. Future large studies utilising deep imaging data from facilities such as Hyper Suprime-Cam and the Large Synoptic Survey Telescope will be ideally placed to further expand upon this work.

\section{Conclusions}

Using a sample of $472$ local Universe ($z<0.06$) galaxies in the stellar mass range $10.25<\log\mathcal{M}_{\star}/\mathcal{M}_{\odot}<10.75$, we explore the variation in galaxy structure as a function of galaxy colour. Our sample of galaxies is sub-divided into red, green and blue colour groups and into elliptical and non-elliptical (disk-type) morphologies. Using KiDS and VIKING derived postage stamp images, a group of eight volunteers visually classify bars, rings, morphological lenses, tidal streams, shells and signs of merger activity for all galaxies. We find a surplus of rings and morphological lenses in disk-type galaxies undergoing transition from the blue cloud to the red sequence. In particular, this surplus appears linked to lenticular and early-type disk galaxies (S0-Sa), i.e., those systems with the most prominent/developed bulges, over other disk-type type systems. The significance of this green surplus relative to the weighted average of the red/blue population is $2.3\sigma$ and $2.9\sigma$ for rings and lenses, respectively. A combined ring plus lens sample shows a $3.3\sigma$ green valley surplus significance. Both the Milky Way and M31, located within the green valley, have previously been shown to exhibit such structural features, in excellent agreement with our findings here. The recovered bar fraction remains flat with colour within disk-type galaxies at $\sim44\%$. Similarly, no detectable trends in recovered tidal streams, shells or merging activity is noted owing to low numbers of these structures being observed, and therefore relatively larger uncertainties.

It is likely that the gradual decline in star formation as galaxies reach the end of their star-forming life and begin to transition from blue to red is the trigger for the trends we observe here. The slow inside-out death of star formation either leads to the construction and subsequent destruction of ring and lens-like structures in disk-type galaxies or, plausibly, the inside-out fading of the disk acts to temporarily reveal the presence of rings and lenses which were previously masked as the galaxy rapidly transitions from blue to red. This therefore implies that the true number fractions of rings and lenses in nature is likely higher than that currently recorded in the literature. Furthermore, whilst no surplus of bars is found as galaxies transition across the green valley, the presence of a bar does act to modulate the presence or otherwise of a ring and, to a lesser extent, a lens. Disk-type galaxies with a bar host $\sim20-30\%$ more rings than those without a bar. Likewise, disk-type galaxies with a bar show a strong surplus of lenses in the green valley, with a surplus significance of $3.0\sigma$. The results of this study strongly support the inside-out death of star formation galaxy evolutionary mechanism, with a slow passive quenching mechanism the driving force. Cataclysmic mechanisms may be ruled out as the dominant source of these trends, as evidenced by the continuing presence of such delicate ring and lens-like structures in the green valley. 

\section*{Acknowledgements}

We thank the referee for their valuable insights which significantly improved the quality of this study.

GAMA is a joint European-Australasian project based around a spectroscopic campaign using the Anglo-Australian Telescope. The GAMA input catalogue is based on data taken from the Sloan Digital Sky Survey and the UKIRT Infrared Deep Sky Survey. Complementary imaging of the GAMA regions is being obtained by a number of independent survey programmes including GALEX MIS, VST KiDS, VISTA VIKING, WISE, Herschel-ATLAS, GMRT and ASKAP providing UV to radio coverage. GAMA is funded by the STFC (UK), the ARC (Australia), the AAO, and the participating institutions. The GAMA website is www.gama-survey.org. 

Based on data products from observations made with ESO Telescopes at the La Silla Paranal Observatory under programme IDs 177.A-3016, 177.A-3017 and 177.A-3018, and on data products produced by Target/OmegaCEN, INAF-OACN, INAF-OAPD and the KiDS production team, on behalf of the KiDS consortium. OmegaCEN and the KiDS production team acknowledge support by NOVA and NWO-M grants. Members of INAF-OAPD and INAF-OACN also acknowledge the support from the Department of Physics \& Astronomy of the University of Padova, and of the Department of Physics of Univ. Federico II (Naples).

Based on observations made with ESO Telescopes at the La Silla Paranal Observatory under programme IDs 179.A-2004. 

This publication uses data generated via the Zooniverse.org platform, development of which is funded by generous support, including a Global Impact Award from Google, and by a grant from the Alfred P. Sloan Foundation.

%%%%%%%%%%%%%%%%%%%%%%%%%%%%%%%%%%%%%%%%%%%%%%%%%%

%%%%%%%%%%%%%%%%%%%% REFERENCES %%%%%%%%%%%%%%%%%%

\bibliographystyle{mnras}
\bibliography{bibtex}

\begin{thebibliography}{}
\makeatletter
\relax
\def\mn@urlcharsother{\let\do\@makeother \do\$\do\&\do\#\do\^\do\_\do\%\do\~}
\def\mn@doi{\begingroup\mn@urlcharsother \@ifnextchar [ {\mn@doi@}
  {\mn@doi@[]}}
\def\mn@doi@[#1]#2{\def\@tempa{#1}\ifx\@tempa\@empty \href
  {http://dx.doi.org/#2} {doi:#2}\else \href {http://dx.doi.org/#2} {#1}\fi
  \endgroup}
\def\mn@eprint#1#2{\mn@eprint@#1:#2::\@nil}
\def\mn@eprint@arXiv#1{\href {http://arxiv.org/abs/#1} {{\tt arXiv:#1}}}
\def\mn@eprint@dblp#1{\href {http://dblp.uni-trier.de/rec/bibtex/#1.xml}
  {dblp:#1}}
\def\mn@eprint@#1:#2:#3:#4\@nil{\def\@tempa {#1}\def\@tempb {#2}\def\@tempc
  {#3}\ifx \@tempc \@empty \let \@tempc \@tempb \let \@tempb \@tempa \fi \ifx
  \@tempb \@empty \def\@tempb {arXiv}\fi \@ifundefined
  {mn@eprint@\@tempb}{\@tempb:\@tempc}{\expandafter \expandafter \csname
  mn@eprint@\@tempb\endcsname \expandafter{\@tempc}}}

\bibitem[\protect\citeauthoryear{{Abazajian} et~al.,}{{Abazajian}
  et~al.}{2009}]{Abazajian2009}
{Abazajian} K.~N.,  et~al., 2009, \mn@doi [\apjs]
  {10.1088/0067-0049/182/2/543}, \href
  {http://adsabs.harvard.edu/abs/2009ApJS..182..543A} {182, 543}

\bibitem[\protect\citeauthoryear{{Agius} et~al.,}{{Agius}
  et~al.}{2013}]{Agius2013}
{Agius} N.~K.,  et~al., 2013, \mn@doi [\mnras] {10.1093/mnras/stt310}, \href
  {http://adsabs.harvard.edu/abs/2013MNRAS.431.1929A} {431, 1929}

\bibitem[\protect\citeauthoryear{{Andrews}, {Kelvin}, {Driver}  \&
  {Robotham}}{{Andrews} et~al.}{2014}]{Andrews2014}
{Andrews} S.~K.,  {Kelvin} L.~S.,  {Driver} S.~P.,   {Robotham} A.~S.~G.,
  2014, \mn@doi [\pasa] {10.1017/pasa.2013.41}, \href
  {http://adsabs.harvard.edu/abs/2014PASA...31....4A} {31, e004}

\bibitem[\protect\citeauthoryear{{Appleton} \& {Struck-Marcell}}{{Appleton} \&
  {Struck-Marcell}}{1996}]{Appleton1996}
{Appleton} P.~N.,  {Struck-Marcell} C.,  1996, \fcp, \href
  {http://adsabs.harvard.edu/abs/1996FCPh...16..111A} {16, 111}

\bibitem[\protect\citeauthoryear{{Arnaboldi}, {Neeser}, {Parker}, {Rosati},
  {Lombardi}, {Dietrich}  \& {Hummel}}{{Arnaboldi}
  et~al.}{2007}]{Arnaboldi2007}
{Arnaboldi} M.,  {Neeser} M.~J.,  {Parker} L.~C.,  {Rosati} P.,  {Lombardi} M.,
   {Dietrich} J.~P.,   {Hummel} W.,  2007, The Messenger, \href
  {http://adsabs.harvard.edu/abs/2007Msngr.127...28A} {127}

\bibitem[\protect\citeauthoryear{{Arp}}{{Arp}}{1966}]{Arp1966}
{Arp} H.,  1966, \mn@doi [\apjs] {10.1086/190147}, \href
  {http://adsabs.harvard.edu/abs/1966ApJS...14....1A} {14, 1}

\bibitem[\protect\citeauthoryear{{Athanassoula} \& {Bosma}}{{Athanassoula} \&
  {Bosma}}{1985}]{Athanassoula1985}
{Athanassoula} E.,  {Bosma} A.,  1985, \mn@doi [\araa]
  {10.1146/annurev.aa.23.090185.001051}, \href
  {http://adsabs.harvard.edu/abs/1985ARA&A..23..147A} {23, 147}

\bibitem[\protect\citeauthoryear{{Athanassoula}, {Machado}  \&
  {Rodionov}}{{Athanassoula} et~al.}{2013}]{Athanassoula2013}
{Athanassoula} E.,  {Machado} R.~E.~G.,   {Rodionov} S.~A.,  2013, \mn@doi
  [\mnras] {10.1093/mnras/sts452}, \href
  {http://adsabs.harvard.edu/abs/2013MNRAS.429.1949A} {429, 1949}

\bibitem[\protect\citeauthoryear{{Baillard} et~al.,}{{Baillard}
  et~al.}{2011}]{Baillard2011}
{Baillard} A.,  et~al., 2011, \mn@doi [\aap] {10.1051/0004-6361/201016423},
  \href {http://adsabs.harvard.edu/abs/2011A&A...532A..74B} {532, A74}

\bibitem[\protect\citeauthoryear{{Baldry}}{{Baldry}}{2008}]{Baldry2008b}
{Baldry} I.~K.,  2008, \mn@doi [Astronomy and Geophysics]
  {10.1111/j.1468-4004.2008.49525.x}, \href
  {http://adsabs.harvard.edu/abs/2008A&G....49e..25B} {49, 5.25}

\bibitem[\protect\citeauthoryear{{Baldry}, {Glazebrook}, {Brinkmann},
  {Ivezi{\'c}}, {Lupton}, {Nichol}  \& {Szalay}}{{Baldry}
  et~al.}{2004}]{Baldry2004}
{Baldry} I.~K.,  {Glazebrook} K.,  {Brinkmann} J.,  {Ivezi{\'c}} {\v Z}.,
  {Lupton} R.~H.,  {Nichol} R.~C.,   {Szalay} A.~S.,  2004, \mn@doi [\apj]
  {10.1086/380092}, \href {http://adsabs.harvard.edu/abs/2004ApJ...600..681B}
  {600, 681}

\bibitem[\protect\citeauthoryear{{Baldry}, {Balogh}, {Bower}, {Glazebrook},
  {Nichol}, {Bamford}  \& {Budavari}}{{Baldry} et~al.}{2006}]{Baldry2006}
{Baldry} I.~K.,  {Balogh} M.~L.,  {Bower} R.~G.,  {Glazebrook} K.,  {Nichol}
  R.~C.,  {Bamford} S.~P.,   {Budavari} T.,  2006, \mn@doi [\mnras]
  {10.1111/j.1365-2966.2006.11081.x}, \href
  {http://adsabs.harvard.edu/abs/2006MNRAS.373..469B} {373, 469}

\bibitem[\protect\citeauthoryear{{Baldry}, {Glazebrook}  \& {Driver}}{{Baldry}
  et~al.}{2008}]{Baldry2008a}
{Baldry} I.~K.,  {Glazebrook} K.,   {Driver} S.~P.,  2008, \mn@doi [\mnras]
  {10.1111/j.1365-2966.2008.13348.x}, \href
  {http://adsabs.harvard.edu/abs/2008MNRAS.388..945B} {388, 945}

\bibitem[\protect\citeauthoryear{{Baldry} et~al.,}{{Baldry}
  et~al.}{2010}]{Baldry2010}
{Baldry} I.~K.,  et~al., 2010, \mn@doi [\mnras]
  {10.1111/j.1365-2966.2010.16282.x}, \href
  {http://adsabs.harvard.edu/abs/2010MNRAS.404...86B} {404, 86}

\bibitem[\protect\citeauthoryear{{Baldry} et~al.,}{{Baldry}
  et~al.}{2012}]{Baldry2012}
{Baldry} I.~K.,  et~al., 2012, \mn@doi [\mnras]
  {10.1111/j.1365-2966.2012.20340.x}, \href
  {http://adsabs.harvard.edu/abs/2012MNRAS.421..621B} {421, 621}

\bibitem[\protect\citeauthoryear{{Baldry} et~al.,}{{Baldry}
  et~al.}{2018}]{Baldry2018}
{Baldry} I.~K.,  et~al., 2018, \mn@doi [\mnras] {10.1093/mnras/stx3042}, \href
  {http://adsabs.harvard.edu/abs/2018MNRAS.474.3875B} {474, 3875}

\bibitem[\protect\citeauthoryear{{Bamford} et~al.,}{{Bamford}
  et~al.}{2009}]{Bamford2009}
{Bamford} S.~P.,  et~al., 2009, \mn@doi [\mnras]
  {10.1111/j.1365-2966.2008.14252.x}, \href
  {http://adsabs.harvard.edu/abs/2009MNRAS.393.1324B} {393, 1324}

\bibitem[\protect\citeauthoryear{{Bland-Hawthorn} \&
  {Gerhard}}{{Bland-Hawthorn} \& {Gerhard}}{2016}]{Bland-Hawthorn2016}
{Bland-Hawthorn} J.,  {Gerhard} O.,  2016, \mn@doi [\araa]
  {10.1146/annurev-astro-081915-023441}, \href
  {http://adsabs.harvard.edu/abs/2016ARA&A..54..529B} {54, 529}

\bibitem[\protect\citeauthoryear{{Block} et~al.,}{{Block}
  et~al.}{2006}]{Block2006}
{Block} D.~L.,  et~al., 2006, \mn@doi [\nat] {10.1038/nature05184}, \href
  {http://adsabs.harvard.edu/abs/2006Natur.443..832B} {443, 832}

\bibitem[\protect\citeauthoryear{{Bremer} et~al.,}{{Bremer}
  et~al.}{2018}]{Bremer2018}
{Bremer} M.~N.,  et~al., 2018, \mn@doi [\mnras] {10.1093/mnras/sty124}, \href
  {http://adsabs.harvard.edu/abs/2018MNRAS.476...12B} {476, 12}

\bibitem[\protect\citeauthoryear{{Bruzual} \& {Charlot}}{{Bruzual} \&
  {Charlot}}{2003}]{Bruzual2003}
{Bruzual} G.,  {Charlot} S.,  2003, \mn@doi [\mnras]
  {10.1046/j.1365-8711.2003.06897.x}, \href
  {http://adsabs.harvard.edu/abs/2003MNRAS.344.1000B} {344, 1000}

\bibitem[\protect\citeauthoryear{{Buta}}{{Buta}}{2013}]{Buta2013}
{Buta} R.~J.,  2013, {Galaxy Morphology}.
p.~1, \mn@doi{10.1007/978-94-007-5609-0_1}

\bibitem[\protect\citeauthoryear{{Buta}}{{Buta}}{2017}]{Buta2017a}
{Buta} R.~J.,  2017, \mn@doi [\mnras] {10.1093/mnras/stx1392}, \href
  {http://adsabs.harvard.edu/abs/2017MNRAS.470.3819B} {470, 3819}

\bibitem[\protect\citeauthoryear{{Buta} \& {Combes}}{{Buta} \&
  {Combes}}{1996}]{Buta1996}
{Buta} R.,  {Combes} F.,  1996, \fcp, \href
  {http://adsabs.harvard.edu/abs/1996FCPh...17...95B} {17, 95}

\bibitem[\protect\citeauthoryear{{Buta}, {Corwin}  \& {Odewahn}}{{Buta}
  et~al.}{2007}]{Buta2007}
{Buta} R.~J.,  {Corwin} H.~G.,   {Odewahn} S.~C.,  2007, {The de Vaucouleurs
  Atlas of Galaxies}.
Cambridge University Press

\bibitem[\protect\citeauthoryear{{Buta} et~al.,}{{Buta}
  et~al.}{2015}]{Buta2015}
{Buta} R.~J.,  et~al., 2015, \mn@doi [\apjs] {10.1088/0067-0049/217/2/32},
  \href {http://adsabs.harvard.edu/abs/2015ApJS..217...32B} {217, 32}

\bibitem[\protect\citeauthoryear{{Calzetti}, {Armus}, {Bohlin}, {Kinney},
  {Koornneef}  \& {Storchi-Bergmann}}{{Calzetti} et~al.}{2000}]{Calzetti2000}
{Calzetti} D.,  {Armus} L.,  {Bohlin} R.~C.,  {Kinney} A.~L.,  {Koornneef} J.,
   {Storchi-Bergmann} T.,  2000, \mn@doi [\apj] {10.1086/308692}, \href
  {http://adsabs.harvard.edu/abs/2000ApJ...533..682C} {533, 682}

\bibitem[\protect\citeauthoryear{{Cameron}}{{Cameron}}{2011}]{Cameron2011}
{Cameron} E.,  2011, \mn@doi [\pasa] {10.1071/AS10046}, \href
  {http://adsabs.harvard.edu/abs/2011PASA...28..128C} {28, 128}

\bibitem[\protect\citeauthoryear{{Capaccioli} \& {Schipani}}{{Capaccioli} \&
  {Schipani}}{2011}]{Capaccioli2011}
{Capaccioli} M.,  {Schipani} P.,  2011, The Messenger, \href
  {http://adsabs.harvard.edu/abs/2011Msngr.146....2C} {146, 2}

\bibitem[\protect\citeauthoryear{{Chabrier}}{{Chabrier}}{2003}]{Chabrier2003}
{Chabrier} G.,  2003, \mn@doi [\pasp] {10.1086/376392}, \href
  {http://adsabs.harvard.edu/abs/2003PASP..115..763C} {115, 763}

\bibitem[\protect\citeauthoryear{{Cluver} et~al.,}{{Cluver}
  et~al.}{2013}]{Cluver2013}
{Cluver} M.~E.,  et~al., 2013, \mn@doi [\apj] {10.1088/0004-637X/765/2/93},
  \href {http://adsabs.harvard.edu/abs/2013ApJ...765...93C} {765, 93}

\bibitem[\protect\citeauthoryear{{Cluver} et~al.,}{{Cluver}
  et~al.}{2014}]{Cluver2014}
{Cluver} M.~E.,  et~al., 2014, \mn@doi [\apj] {10.1088/0004-637X/782/2/90},
  \href {http://adsabs.harvard.edu/abs/2014ApJ...782...90C} {782, 90}

\bibitem[\protect\citeauthoryear{{Comer{\'o}n}, {Knapen}, {Beckman},
  {Laurikainen}, {Salo}, {Mart{\'{\i}}nez-Valpuesta}  \& {Buta}}{{Comer{\'o}n}
  et~al.}{2010}]{Comeron2010}
{Comer{\'o}n} S.,  {Knapen} J.~H.,  {Beckman} J.~E.,  {Laurikainen} E.,  {Salo}
  H.,  {Mart{\'{\i}}nez-Valpuesta} I.,   {Buta} R.~J.,  2010, \mn@doi [\mnras]
  {10.1111/j.1365-2966.2009.16057.x}, \href
  {http://adsabs.harvard.edu/abs/2010MNRAS.402.2462C} {402, 2462}

\bibitem[\protect\citeauthoryear{{Comer{\'o}n} et~al.,}{{Comer{\'o}n}
  et~al.}{2014}]{Comeron2014}
{Comer{\'o}n} S.,  et~al., 2014, \mn@doi [\aap] {10.1051/0004-6361/201321633},
  \href {http://adsabs.harvard.edu/abs/2014A&A...562A.121C} {562, A121}

\bibitem[\protect\citeauthoryear{{Cullen}, {Alexander}, {Green}  \&
  {Sheth}}{{Cullen} et~al.}{2007}]{Cullen2007}
{Cullen} H.,  {Alexander} P.,  {Green} D.~A.,   {Sheth} K.,  2007, \mn@doi
  [\mnras] {10.1111/j.1365-2966.2007.11506.x}, \href
  {http://adsabs.harvard.edu/abs/2007MNRAS.376...98C} {376, 98}

\bibitem[\protect\citeauthoryear{{Dalton}, {Sutherland}, {Emerson},
  {Woodhouse}, {Terrett}  \& {Whalley}}{{Dalton} et~al.}{2010}]{Dalton2010}
{Dalton} G.~B.,  {Sutherland} W.~J.,  {Emerson} J.~P.,  {Woodhouse} G.~F.~W.,
  {Terrett} D.~L.,   {Whalley} M.~S.,  2010, in Ground-based and Airborne
  Instrumentation for Astronomy III. p. 77351J, \mn@doi{10.1117/12.857186}

\bibitem[\protect\citeauthoryear{{De Propris} et~al.,}{{De Propris}
  et~al.}{2014}]{DePropris2014}
{De Propris} R.,  et~al., 2014, \mn@doi [\mnras] {10.1093/mnras/stu1452}, \href
  {http://adsabs.harvard.edu/abs/2014MNRAS.444.2200D} {444, 2200}

\bibitem[\protect\citeauthoryear{{Diaferio}, {Kauffmann}, {Balogh}, {White},
  {Schade}  \& {Ellingson}}{{Diaferio} et~al.}{2001}]{Diaferio2001}
{Diaferio} A.,  {Kauffmann} G.,  {Balogh} M.~L.,  {White} S.~D.~M.,  {Schade}
  D.,   {Ellingson} E.,  2001, \mn@doi [\mnras]
  {10.1046/j.1365-8711.2001.04303.x}, \href
  {http://adsabs.harvard.edu/abs/2001MNRAS.323..999D} {323, 999}

\bibitem[\protect\citeauthoryear{{Dressler}}{{Dressler}}{1980}]{Dressler1980}
{Dressler} A.,  1980, \mn@doi [\apj] {10.1086/157753}, \href
  {http://adsabs.harvard.edu/abs/1980ApJ...236..351D} {236, 351}

\bibitem[\protect\citeauthoryear{{Driver} et~al.,}{{Driver}
  et~al.}{2006}]{Driver2006}
{Driver} S.~P.,  et~al., 2006, \mn@doi [\mnras]
  {10.1111/j.1365-2966.2006.10126.x}, \href
  {http://adsabs.harvard.edu/abs/2006MNRAS.368..414D} {368, 414}

\bibitem[\protect\citeauthoryear{{Driver} et~al.,}{{Driver}
  et~al.}{2009}]{Driver2009}
{Driver} S.~P.,  et~al., 2009, \mn@doi [Astronomy and Geophysics]
  {10.1111/j.1468-4004.2009.50512.x}, \href
  {http://adsabs.harvard.edu/abs/2009A&G....50e..12D} {50, 5.12}

\bibitem[\protect\citeauthoryear{{Driver} et~al.,}{{Driver}
  et~al.}{2016}]{Driver2016}
{Driver} S.~P.,  et~al., 2016, \mn@doi [\mnras] {10.1093/mnras/stv2505}, \href
  {http://adsabs.harvard.edu/abs/2016MNRAS.455.3911D} {455, 3911}

\bibitem[\protect\citeauthoryear{{Dye} et~al.,}{{Dye} et~al.}{2006}]{Dye2006}
{Dye} S.,  et~al., 2006, \mn@doi [\mnras] {10.1111/j.1365-2966.2006.10928.x},
  \href {http://adsabs.harvard.edu/abs/2006MNRAS.372.1227D} {372, 1227}

\bibitem[\protect\citeauthoryear{{Eales}, {de Vis}, {Smith}, {Appah}, {Ciesla},
  {Duffield}  \& {Schofield}}{{Eales} et~al.}{2017}]{Eales2017}
{Eales} S.,  {de Vis} P.,  {Smith} M.~W.~L.,  {Appah} K.,  {Ciesla} L.,
  {Duffield} C.,   {Schofield} S.,  2017, \mn@doi [\mnras]
  {10.1093/mnras/stw2875}, \href
  {http://adsabs.harvard.edu/abs/2017MNRAS.465.3125E} {465, 3125}

\bibitem[\protect\citeauthoryear{{Eales} et~al.,}{{Eales}
  et~al.}{2018}]{Eales2018}
{Eales} S.,  et~al., 2018, \mn@doi [\mnras] {10.1093/mnras/stx2548}, \href
  {http://adsabs.harvard.edu/abs/2018MNRAS.473.3507E} {473, 3507}

\bibitem[\protect\citeauthoryear{{Edge}, {Sutherland}, {Kuijken}, {Driver},
  {McMahon}, {Eales}  \& {Emerson}}{{Edge} et~al.}{2013}]{Edge2013}
{Edge} A.,  {Sutherland} W.,  {Kuijken} K.,  {Driver} S.,  {McMahon} R.,
  {Eales} S.,   {Emerson} J.~P.,  2013, The Messenger, \href
  {http://adsabs.harvard.edu/abs/2013Msngr.154...32E} {154, 32}

\bibitem[\protect\citeauthoryear{{Elmegreen} \& {Elmegreen}}{{Elmegreen} \&
  {Elmegreen}}{1985}]{Elmegreen1985}
{Elmegreen} B.~G.,  {Elmegreen} D.~M.,  1985, \mn@doi [\apj] {10.1086/162810},
  \href {http://adsabs.harvard.edu/abs/1985ApJ...288..438E} {288, 438}

\bibitem[\protect\citeauthoryear{{Erdo{\v g}du} et~al.,}{{Erdo{\v g}du}
  et~al.}{2006}]{Erdogdu2006}
{Erdo{\v g}du} P.,  et~al., 2006, \mn@doi [\mnras]
  {10.1111/j.1365-2966.2006.10243.x}, \href
  {http://adsabs.harvard.edu/abs/2006MNRAS.368.1515E} {368, 1515}

\bibitem[\protect\citeauthoryear{{Fang}, {Faber}, {Salim}, {Graves}  \&
  {Rich}}{{Fang} et~al.}{2012}]{Fang2012}
{Fang} J.~J.,  {Faber} S.~M.,  {Salim} S.,  {Graves} G.~J.,   {Rich} R.~M.,
  2012, \mn@doi [\apj] {10.1088/0004-637X/761/1/23}, \href
  {http://adsabs.harvard.edu/abs/2012ApJ...761...23F} {761, 23}

\bibitem[\protect\citeauthoryear{{Fang}, {Faber}, {Koo}  \& {Dekel}}{{Fang}
  et~al.}{2013}]{Fang2013}
{Fang} J.~J.,  {Faber} S.~M.,  {Koo} D.~C.,   {Dekel} A.,  2013, \mn@doi [\apj]
  {10.1088/0004-637X/776/1/63}, \href
  {http://adsabs.harvard.edu/abs/2013ApJ...776...63F} {776, 63}

\bibitem[\protect\citeauthoryear{{Feldmann}}{{Feldmann}}{2017}]{Feldmann2017b}
{Feldmann} R.,  2017, \mn@doi [\mnras] {10.1093/mnrasl/slx073}, \href
  {http://adsabs.harvard.edu/abs/2017MNRAS.470L..59F} {470, L59}

\bibitem[\protect\citeauthoryear{{Feldmann}, {Quataert}, {Hopkins},
  {Faucher-Gigu{\`e}re}  \& {Kere{\v s}}}{{Feldmann}
  et~al.}{2017}]{Feldmann2017a}
{Feldmann} R.,  {Quataert} E.,  {Hopkins} P.~F.,  {Faucher-Gigu{\`e}re} C.-A.,
   {Kere{\v s}} D.,  2017, \mn@doi [\mnras] {10.1093/mnras/stx1120}, \href
  {http://adsabs.harvard.edu/abs/2017MNRAS.470.1050F} {470, 1050}

\bibitem[\protect\citeauthoryear{{Freeman}}{{Freeman}}{1975}]{Freeman1975}
{Freeman} K.~C.,  1975, in {Hayli} A.,  ed.,  IAU Symposium Vol. 69, Dynamics
  of the Solar Systems. p.~367

\bibitem[\protect\citeauthoryear{{Graham} \& {Driver}}{{Graham} \&
  {Driver}}{2005}]{Graham2005a}
{Graham} A.~W.,  {Driver} S.~P.,  2005, \mn@doi [\pasa] {10.1071/AS05001},
  \href {http://adsabs.harvard.edu/abs/2005PASA...22..118G} {22, 118}

\bibitem[\protect\citeauthoryear{{Gunn} \& {Gott}}{{Gunn} \&
  {Gott}}{1972}]{Gunn1972}
{Gunn} J.~E.,  {Gott} III J.~R.,  1972, \mn@doi [\apj] {10.1086/151605}, \href
  {http://adsabs.harvard.edu/abs/1972ApJ...176....1G} {176, 1}

\bibitem[\protect\citeauthoryear{{Hart} et~al.,}{{Hart}
  et~al.}{2016}]{Hart2016}
{Hart} R.~E.,  et~al., 2016, \mn@doi [\mnras] {10.1093/mnras/stw1588}, \href
  {http://adsabs.harvard.edu/abs/2016MNRAS.461.3663H} {461, 3663}

\bibitem[\protect\citeauthoryear{{Hiemer}, {Barden}, {Kelvin},
  {H{\"a}u{\ss}ler}  \& {Schindler}}{{Hiemer} et~al.}{2014}]{Hiemer2014}
{Hiemer} A.,  {Barden} M.,  {Kelvin} L.~S.,  {H{\"a}u{\ss}ler} B.,
  {Schindler} S.,  2014, \mn@doi [\mnras] {10.1093/mnras/stu1649}, \href
  {http://adsabs.harvard.edu/abs/2014MNRAS.444.3089H} {444, 3089}

\bibitem[\protect\citeauthoryear{{Hill} et~al.,}{{Hill}
  et~al.}{2011}]{Hill2011}
{Hill} D.~T.,  et~al., 2011, \mn@doi [\mnras]
  {10.1111/j.1365-2966.2010.17950.x}, \href
  {http://adsabs.harvard.edu/abs/2011MNRAS.412..765H} {412, 765}

\bibitem[\protect\citeauthoryear{{Hubble}}{{Hubble}}{1926}]{Hubble1926}
{Hubble} E.~P.,  1926, \mn@doi [\apj] {10.1086/143018}, \href
  {http://adsabs.harvard.edu/abs/1926ApJ....64..321H} {64}

\bibitem[\protect\citeauthoryear{{Hummel}, {Hanuschik}, {de Bilbao}, {Mieske},
  {Szeifert}, {Ivanov}  \& {Castro}}{{Hummel} et~al.}{2010}]{Hummel2010}
{Hummel} W.,  {Hanuschik} R.,  {de Bilbao} L.,  {Mieske} S.,  {Szeifert} T.,
  {Ivanov} V.,   {Castro} S.,  2010, in Observatory Operations: Strategies,
  Processes, and Systems III. p. 77371H, \mn@doi{10.1117/12.856414}

\bibitem[\protect\citeauthoryear{{James} \& {Percival}}{{James} \&
  {Percival}}{2015}]{James2015}
{James} P.~A.,  {Percival} S.~M.,  2015, \mn@doi [\mnras]
  {10.1093/mnras/stv846}, \href
  {http://adsabs.harvard.edu/abs/2015MNRAS.450.3503J} {450, 3503}

\bibitem[\protect\citeauthoryear{{James} \& {Percival}}{{James} \&
  {Percival}}{2016}]{James2016}
{James} P.~A.,  {Percival} S.~M.,  2016, \mn@doi [\mnras]
  {10.1093/mnras/stv2978}, \href
  {http://adsabs.harvard.edu/abs/2016MNRAS.457..917J} {457, 917}

\bibitem[\protect\citeauthoryear{{James} \& {Percival}}{{James} \&
  {Percival}}{2018}]{James2018}
{James} P.~A.,  {Percival} S.~M.,  2018, \mn@doi [\mnras]
  {10.1093/mnras/stx2990}, \href
  {http://adsabs.harvard.edu/abs/2018MNRAS.474.3101J} {474, 3101}

\bibitem[\protect\citeauthoryear{{Jeans}}{{Jeans}}{1919}]{Jeans1919}
{Jeans} J.~H.,  1919, {Problems of cosmogony and stellar dynamics}

\bibitem[\protect\citeauthoryear{{Kauffmann}, {White}  \&
  {Guiderdoni}}{{Kauffmann} et~al.}{1993}]{Kauffmann1993}
{Kauffmann} G.,  {White} S.~D.~M.,   {Guiderdoni} B.,  1993, \mn@doi [\mnras]
  {10.1093/mnras/264.1.201}, \href
  {http://adsabs.harvard.edu/abs/1993MNRAS.264..201K} {264, 201}

\bibitem[\protect\citeauthoryear{{Kelvin} et~al.,}{{Kelvin}
  et~al.}{2012}]{Kelvin2012}
{Kelvin} L.~S.,  et~al., 2012, \mn@doi [\mnras]
  {10.1111/j.1365-2966.2012.20355.x}, \href
  {http://adsabs.harvard.edu/abs/2012MNRAS.421.1007K} {421, 1007}

\bibitem[\protect\citeauthoryear{{Kelvin} et~al.,}{{Kelvin}
  et~al.}{2014a}]{Kelvin2014a}
{Kelvin} L.~S.,  et~al., 2014a, \mn@doi [\mnras] {10.1093/mnras/stt2391}, \href
  {http://adsabs.harvard.edu/abs/2014MNRAS.439.1245K} {439, 1245}

\bibitem[\protect\citeauthoryear{{Kelvin} et~al.,}{{Kelvin}
  et~al.}{2014b}]{Kelvin2014b}
{Kelvin} L.~S.,  et~al., 2014b, \mn@doi [\mnras] {10.1093/mnras/stu1507}, \href
  {http://adsabs.harvard.edu/abs/2014MNRAS.444.1647K} {444, 1647}

\bibitem[\protect\citeauthoryear{{Kettlety} et~al.,}{{Kettlety}
  et~al.}{2018}]{Kettlety2018}
{Kettlety} T.,  et~al., 2018, \mn@doi [\mnras] {10.1093/mnras/stx2379}, \href
  {http://adsabs.harvard.edu/abs/2018MNRAS.473..776K} {473, 776}

\bibitem[\protect\citeauthoryear{{Knapen}}{{Knapen}}{2010}]{Knapen2010}
{Knapen} J.~H.,  2010, in {Block} D.~L.,  {Freeman} K.~C.,   {Puerari} I.,
  eds, Galaxies and their Masks. p.~201 (\mn@eprint {arXiv} {1005.0506}),
  \mn@doi{10.1007/978-1-4419-7317-7_18}

\bibitem[\protect\citeauthoryear{{Knapen} \& {James}}{{Knapen} \&
  {James}}{2009}]{Knapen2009}
{Knapen} J.~H.,  {James} P.~A.,  2009, \mn@doi [\apj]
  {10.1088/0004-637X/698/2/1437}, \href
  {http://adsabs.harvard.edu/abs/2009ApJ...698.1437K} {698, 1437}

\bibitem[\protect\citeauthoryear{{Kormendy}}{{Kormendy}}{1979}]{Kormendy1979}
{Kormendy} J.,  1979, \mn@doi [\apj] {10.1086/156782}, \href
  {http://adsabs.harvard.edu/abs/1979ApJ...227..714K} {227, 714}

\bibitem[\protect\citeauthoryear{{Kormendy} \& {Djorgovski}}{{Kormendy} \&
  {Djorgovski}}{1989}]{Kormendy1989}
{Kormendy} J.,  {Djorgovski} S.,  1989, \mn@doi [\araa]
  {10.1146/annurev.aa.27.090189.001315}, \href
  {http://adsabs.harvard.edu/abs/1989ARA&A..27..235K} {27, 235}

\bibitem[\protect\citeauthoryear{{Kruk} et~al.,}{{Kruk}
  et~al.}{2018}]{Kruk2018}
{Kruk} S.~J.,  et~al., 2018, \mn@doi [\mnras] {10.1093/mnras/stx2605}, \href
  {http://adsabs.harvard.edu/abs/2018MNRAS.473.4731K} {473, 4731}

\bibitem[\protect\citeauthoryear{{Kuijken}}{{Kuijken}}{2011}]{Kuijken2011}
{Kuijken} K.,  2011, The Messenger, \href
  {http://adsabs.harvard.edu/abs/2011Msngr.146....8K} {146, 8}

\bibitem[\protect\citeauthoryear{{Larson}, {Tinsley}  \& {Caldwell}}{{Larson}
  et~al.}{1980}]{Larson1980}
{Larson} R.~B.,  {Tinsley} B.~M.,   {Caldwell} C.~N.,  1980, \mn@doi [\apj]
  {10.1086/157917}, \href {http://adsabs.harvard.edu/abs/1980ApJ...237..692L}
  {237, 692}

\bibitem[\protect\citeauthoryear{{Laurikainen} \& {Salo}}{{Laurikainen} \&
  {Salo}}{2017}]{Laurikainen2017}
{Laurikainen} E.,  {Salo} H.,  2017, \mn@doi [\aap]
  {10.1051/0004-6361/201628936}, \href
  {http://adsabs.harvard.edu/abs/2017A&A...598A..10L} {598, A10}

\bibitem[\protect\citeauthoryear{{Laurikainen}, {Salo}, {Athanassoula},
  {Bosma}, {Buta}  \& {Janz}}{{Laurikainen} et~al.}{2013}]{Laurikainen2013}
{Laurikainen} E.,  {Salo} H.,  {Athanassoula} E.,  {Bosma} A.,  {Buta} R.,
  {Janz} J.,  2013, \mn@doi [\mnras] {10.1093/mnras/stt150}, \href
  {http://adsabs.harvard.edu/abs/2013MNRAS.430.3489L} {430, 3489}

\bibitem[\protect\citeauthoryear{{Lawrence} et~al.,}{{Lawrence}
  et~al.}{2007}]{Lawrence2007}
{Lawrence} A.,  et~al., 2007, \mn@doi [\mnras]
  {10.1111/j.1365-2966.2007.12040.x}, \href
  {http://adsabs.harvard.edu/abs/2007MNRAS.379.1599L} {379, 1599}

\bibitem[\protect\citeauthoryear{{Lintott} et~al.,}{{Lintott}
  et~al.}{2008}]{Lintott2008}
{Lintott} C.~J.,  et~al., 2008, \mn@doi [\mnras]
  {10.1111/j.1365-2966.2008.13689.x}, \href
  {http://adsabs.harvard.edu/abs/2008MNRAS.389.1179L} {389, 1179}

\bibitem[\protect\citeauthoryear{{Liske} et~al.,}{{Liske}
  et~al.}{2015}]{Liske2015}
{Liske} J.,  et~al., 2015, \mn@doi [\mnras] {10.1093/mnras/stv1436}, \href
  {http://adsabs.harvard.edu/abs/2015MNRAS.452.2087L} {452, 2087}

\bibitem[\protect\citeauthoryear{{Lupton}, {Blanton}, {Fekete}, {Hogg},
  {O'Mullane}, {Szalay}  \& {Wherry}}{{Lupton} et~al.}{2004}]{Lupton2004}
{Lupton} R.,  {Blanton} M.~R.,  {Fekete} G.,  {Hogg} D.~W.,  {O'Mullane} W.,
  {Szalay} A.,   {Wherry} N.,  2004, \mn@doi [\pasp] {10.1086/382245}, \href
  {http://adsabs.harvard.edu/abs/2004PASP..116..133L} {116, 133}

\bibitem[\protect\citeauthoryear{{Malin} \& {Carter}}{{Malin} \&
  {Carter}}{1980}]{Malin1980}
{Malin} D.~F.,  {Carter} D.,  1980, \mn@doi [\nat] {10.1038/285643a0}, \href
  {http://adsabs.harvard.edu/abs/1980Natur.285..643M} {285, 643}

\bibitem[\protect\citeauthoryear{{Malin} \& {Carter}}{{Malin} \&
  {Carter}}{1983}]{Malin1983}
{Malin} D.~F.,  {Carter} D.,  1983, \mn@doi [\apj] {10.1086/161467}, \href
  {http://adsabs.harvard.edu/abs/1983ApJ...274..534M} {274, 534}

\bibitem[\protect\citeauthoryear{{Martig}, {Bournaud}, {Teyssier}  \&
  {Dekel}}{{Martig} et~al.}{2009}]{Martig2009}
{Martig} M.,  {Bournaud} F.,  {Teyssier} R.,   {Dekel} A.,  2009, \mn@doi
  [\apj] {10.1088/0004-637X/707/1/250}, \href
  {http://adsabs.harvard.edu/abs/2009ApJ...707..250M} {707, 250}

\bibitem[\protect\citeauthoryear{{Martin} et~al.,}{{Martin}
  et~al.}{2007}]{Martin2007}
{Martin} D.~C.,  et~al., 2007, \mn@doi [\apjs] {10.1086/516639}, \href
  {http://adsabs.harvard.edu/abs/2007ApJS..173..342M} {173, 342}

\bibitem[\protect\citeauthoryear{{Mart{\'{\i}}nez-Delgado}
  et~al.,}{{Mart{\'{\i}}nez-Delgado} et~al.}{2010}]{Martinez-Delgado2010}
{Mart{\'{\i}}nez-Delgado} D.,  et~al., 2010, \mn@doi [\aj]
  {10.1088/0004-6256/140/4/962}, \href
  {http://adsabs.harvard.edu/abs/2010AJ....140..962M} {140, 962}

\bibitem[\protect\citeauthoryear{{Mart{\'{\i}}nez-Delgado}, {D'Onghia},
  {Chonis}, {Beaton}, {Teuwen}, {GaBany}, {Grebel}  \&
  {Morales}}{{Mart{\'{\i}}nez-Delgado} et~al.}{2015}]{Martinez-Delgado2015}
{Mart{\'{\i}}nez-Delgado} D.,  {D'Onghia} E.,  {Chonis} T.~S.,  {Beaton} R.~L.,
   {Teuwen} K.,  {GaBany} R.~J.,  {Grebel} E.~K.,   {Morales} G.,  2015,
  \mn@doi [\aj] {10.1088/0004-6256/150/4/116}, \href
  {http://adsabs.harvard.edu/abs/2015AJ....150..116M} {150, 116}

\bibitem[\protect\citeauthoryear{{Masters} et~al.,}{{Masters}
  et~al.}{2010}]{Masters2010}
{Masters} K.~L.,  et~al., 2010, \mn@doi [\mnras]
  {10.1111/j.1365-2966.2010.16503.x}, \href
  {http://adsabs.harvard.edu/abs/2010MNRAS.405..783M} {405, 783}

\bibitem[\protect\citeauthoryear{{Mendez}, {Coil}, {Lotz}, {Salim}, {Moustakas}
   \& {Simard}}{{Mendez} et~al.}{2011}]{Mendez2011}
{Mendez} A.~J.,  {Coil} A.~L.,  {Lotz} J.,  {Salim} S.,  {Moustakas} J.,
  {Simard} L.,  2011, \mn@doi [\apj] {10.1088/0004-637X/736/2/110}, \href
  {http://adsabs.harvard.edu/abs/2011ApJ...736..110M} {736, 110}

\bibitem[\protect\citeauthoryear{{Moffett} et~al.,}{{Moffett}
  et~al.}{2016}]{Moffett2016a}
{Moffett} A.~J.,  et~al., 2016, \mn@doi [\mnras] {10.1093/mnras/stv2883}, \href
  {http://adsabs.harvard.edu/abs/2016MNRAS.457.1308M} {457, 1308}

\bibitem[\protect\citeauthoryear{{Moore}, {Katz}, {Lake}, {Dressler}  \&
  {Oemler}}{{Moore} et~al.}{1996}]{Moore1996}
{Moore} B.,  {Katz} N.,  {Lake} G.,  {Dressler} A.,   {Oemler} A.,  1996,
  \mn@doi [\nat] {10.1038/379613a0}, \href
  {http://adsabs.harvard.edu/abs/1996Natur.379..613M} {379, 613}

\bibitem[\protect\citeauthoryear{{Mutch}, {Croton}  \& {Poole}}{{Mutch}
  et~al.}{2011}]{Mutch2011}
{Mutch} S.~J.,  {Croton} D.~J.,   {Poole} G.~B.,  2011, \mn@doi [\apj]
  {10.1088/0004-637X/736/2/84}, \href
  {http://adsabs.harvard.edu/abs/2011ApJ...736...84M} {736, 84}

\bibitem[\protect\citeauthoryear{{Oemler}, {Abramson}, {Gladders}, {Dressler},
  {Poggianti}  \& {Vulcani}}{{Oemler} et~al.}{2017}]{Oemler2017}
{Oemler} Jr. A.,  {Abramson} L.~E.,  {Gladders} M.~D.,  {Dressler} A.,
  {Poggianti} B.~M.,   {Vulcani} B.,  2017, \mn@doi [\apj]
  {10.3847/1538-4357/aa789e}, \href
  {http://adsabs.harvard.edu/abs/2017ApJ...844...45O} {844, 45}

\bibitem[\protect\citeauthoryear{{Pagani}, {Lequeux}, {Cesarsky}, {Donas},
  {Milliard}, {Loinard}  \& {Sauvage}}{{Pagani} et~al.}{1999}]{Pagani1999}
{Pagani} L.,  {Lequeux} J.,  {Cesarsky} D.,  {Donas} J.,  {Milliard} B.,
  {Loinard} L.,   {Sauvage} M.,  1999, \aap, \href
  {http://adsabs.harvard.edu/abs/1999A&A...351..447P} {351, 447}

\bibitem[\protect\citeauthoryear{{Park}, {Gott}  \& {Choi}}{{Park}
  et~al.}{2008}]{Park2008}
{Park} C.,  {Gott} III J.~R.,   {Choi} Y.-Y.,  2008, \mn@doi [\apj]
  {10.1086/524192}, \href {http://adsabs.harvard.edu/abs/2008ApJ...674..784P}
  {674, 784}

\bibitem[\protect\citeauthoryear{{Peng} et~al.,}{{Peng}
  et~al.}{2010}]{Peng2010b}
{Peng} Y.-j.,  et~al., 2010, \mn@doi [\apj] {10.1088/0004-637X/721/1/193},
  \href {http://adsabs.harvard.edu/abs/2010ApJ...721..193P} {721, 193}

\bibitem[\protect\citeauthoryear{{Peng}, {Lilly}, {Renzini}  \&
  {Carollo}}{{Peng} et~al.}{2012}]{Peng2012}
{Peng} Y.-j.,  {Lilly} S.~J.,  {Renzini} A.,   {Carollo} M.,  2012, \mn@doi
  [\apj] {10.1088/0004-637X/757/1/4}, \href
  {http://adsabs.harvard.edu/abs/2012ApJ...757....4P} {757, 4}

\bibitem[\protect\citeauthoryear{{P{\'e}rez} et~al.,}{{P{\'e}rez}
  et~al.}{2013}]{Perez2013}
{P{\'e}rez} E.,  et~al., 2013, \mn@doi [\apjl] {10.1088/2041-8205/764/1/L1},
  \href {http://adsabs.harvard.edu/abs/2013ApJ...764L...1P} {764, L1}

\bibitem[\protect\citeauthoryear{{Perrine}}{{Perrine}}{1922}]{Perrine1922}
{Perrine} C.~D.,  1922, \mn@doi [\mnras] {10.1093/mnras/82.8.486}, \href
  {http://adsabs.harvard.edu/abs/1922MNRAS..82..486P} {82, 486}

\bibitem[\protect\citeauthoryear{{Reynolds}}{{Reynolds}}{1920}]{Reynolds1920}
{Reynolds} J.~H.,  1920, \mn@doi [\mnras] {10.1093/mnras/80.8.746}, \href
  {http://adsabs.harvard.edu/abs/1920MNRAS..80..746R} {80, 746}

\bibitem[\protect\citeauthoryear{{Robotham} et~al.,}{{Robotham}
  et~al.}{2011}]{Robotham2011b}
{Robotham} A.~S.~G.,  et~al., 2011, \mn@doi [\mnras]
  {10.1111/j.1365-2966.2011.19217.x}, \href
  {http://adsabs.harvard.edu/abs/2011MNRAS.416.2640R} {416, 2640}

\bibitem[\protect\citeauthoryear{{Rowlands} et~al.,}{{Rowlands}
  et~al.}{2018}]{Rowlands2018}
{Rowlands} K.,  et~al., 2018, \mn@doi [\mnras] {10.1093/mnras/stx1903}, \href
  {http://adsabs.harvard.edu/abs/2018MNRAS.473.1168R} {473, 1168}

\bibitem[\protect\citeauthoryear{{Saito}, {Zoccali}, {McWilliam}, {Minniti},
  {Gonzalez}  \& {Hill}}{{Saito} et~al.}{2011}]{Saito2011}
{Saito} R.~K.,  {Zoccali} M.,  {McWilliam} A.,  {Minniti} D.,  {Gonzalez}
  O.~A.,   {Hill} V.,  2011, \mn@doi [\aj] {10.1088/0004-6256/142/3/76}, \href
  {http://adsabs.harvard.edu/abs/2011AJ....142...76S} {142, 76}

\bibitem[\protect\citeauthoryear{{Salo} \& {Laurikainen}}{{Salo} \&
  {Laurikainen}}{2000a}]{Salo2000a}
{Salo} H.,  {Laurikainen} E.,  2000a, \mn@doi [\mnras]
  {10.1046/j.1365-8711.2000.03650.x}, \href
  {http://adsabs.harvard.edu/abs/2000MNRAS.319..377S} {319, 377}

\bibitem[\protect\citeauthoryear{{Salo} \& {Laurikainen}}{{Salo} \&
  {Laurikainen}}{2000b}]{Salo2000b}
{Salo} H.,  {Laurikainen} E.,  2000b, \mn@doi [\mnras]
  {10.1046/j.1365-8711.2000.03651.x}, \href
  {http://adsabs.harvard.edu/abs/2000MNRAS.319..393S} {319, 393}

\bibitem[\protect\citeauthoryear{{Salo} \& {Laurikainen}}{{Salo} \&
  {Laurikainen}}{2017}]{Salo2017}
{Salo} H.,  {Laurikainen} E.,  2017, \mn@doi [\apj]
  {10.3847/1538-4357/835/2/252}, \href
  {http://adsabs.harvard.edu/abs/2017ApJ...835..252S} {835, 252}

\bibitem[\protect\citeauthoryear{{Sandage}}{{Sandage}}{1961}]{Sandage1961}
{Sandage} A.,  1961, {The Hubble Atlas of Galaxies}

\bibitem[\protect\citeauthoryear{{Sandage}}{{Sandage}}{2005}]{Sandage2005}
{Sandage} A.,  2005, \mn@doi [\araa] {10.1146/annurev.astro.43.112904.104839},
  \href {http://adsabs.harvard.edu/abs/2005ARA&A..43..581S} {43, 581}

\bibitem[\protect\citeauthoryear{{Schawinski} et~al.,}{{Schawinski}
  et~al.}{2014}]{Schawinski2014}
{Schawinski} K.,  et~al., 2014, \mn@doi [\mnras] {10.1093/mnras/stu327}, \href
  {http://adsabs.harvard.edu/abs/2014MNRAS.440..889S} {440, 889}

\bibitem[\protect\citeauthoryear{{Schiminovich} et~al.,}{{Schiminovich}
  et~al.}{2007}]{Schiminovich2007}
{Schiminovich} D.,  et~al., 2007, \mn@doi [\apjs] {10.1086/524659}, \href
  {http://adsabs.harvard.edu/abs/2007ApJS..173..315S} {173, 315}

\bibitem[\protect\citeauthoryear{{Schwarz}}{{Schwarz}}{1981}]{Schwarz1981}
{Schwarz} M.~P.,  1981, \mn@doi [\apj] {10.1086/159011}, \href
  {http://adsabs.harvard.edu/abs/1981ApJ...247...77S} {247, 77}

\bibitem[\protect\citeauthoryear{{Schwarz}}{{Schwarz}}{1984}]{Schwarz1984}
{Schwarz} M.~P.,  1984, \mn@doi [\mnras] {10.1093/mnras/209.1.93}, \href
  {http://adsabs.harvard.edu/abs/1984MNRAS.209...93S} {209, 93}

\bibitem[\protect\citeauthoryear{{Schweizer} \& {Seitzer}}{{Schweizer} \&
  {Seitzer}}{1988}]{Schweizer1988}
{Schweizer} F.,  {Seitzer} P.,  1988, \mn@doi [\apj] {10.1086/166270}, \href
  {http://adsabs.harvard.edu/abs/1988ApJ...328...88S} {328, 88}

\bibitem[\protect\citeauthoryear{{Schweizer}, {Ford}, {Jedrzejewski}  \&
  {Giovanelli}}{{Schweizer} et~al.}{1987}]{Schweizer1987}
{Schweizer} F.,  {Ford} Jr. W.~K.,  {Jedrzejewski} R.,   {Giovanelli} R.,
  1987, \mn@doi [\apj] {10.1086/165562}, \href
  {http://adsabs.harvard.edu/abs/1987ApJ...320..454S} {320, 454}

\bibitem[\protect\citeauthoryear{{S{\'e}rsic}}{{S{\'e}rsic}}{1963}]{Sersic1963}
{S{\'e}rsic} J.~L.,  1963, Boletin de la Asociacion Argentina de Astronomia La
  Plata Argentina, \href {http://adsabs.harvard.edu/abs/1963BAAA....6...41S}
  {6, 41}

\bibitem[\protect\citeauthoryear{{S{\'e}rsic}}{{S{\'e}rsic}}{1968}]{Sersic1968}
{S{\'e}rsic} J.~L.,  1968, {Atlas de Galaxias Australes}

\bibitem[\protect\citeauthoryear{{Sheth} et~al.,}{{Sheth}
  et~al.}{2008}]{Sheth2008}
{Sheth} K.,  et~al., 2008, \mn@doi [\apj] {10.1086/524980}, \href
  {http://adsabs.harvard.edu/abs/2008ApJ...675.1141S} {675, 1141}

\bibitem[\protect\citeauthoryear{{Simard}, {Mendel}, {Patton}, {Ellison}  \&
  {McConnachie}}{{Simard} et~al.}{2011}]{Simard2011}
{Simard} L.,  {Mendel} J.~T.,  {Patton} D.~R.,  {Ellison} S.~L.,
  {McConnachie} A.~W.,  2011, \mn@doi [\apjs] {10.1088/0067-0049/196/1/11},
  \href {http://adsabs.harvard.edu/abs/2011ApJS..196...11S} {196, 11}

\bibitem[\protect\citeauthoryear{{Smethurst} et~al.,}{{Smethurst}
  et~al.}{2015}]{Smethurst2015}
{Smethurst} R.~J.,  et~al., 2015, \mn@doi [\mnras] {10.1093/mnras/stv161},
  \href {http://adsabs.harvard.edu/abs/2015MNRAS.450..435S} {450, 435}

\bibitem[\protect\citeauthoryear{{Strateva} et~al.,}{{Strateva}
  et~al.}{2001}]{Strateva2001}
{Strateva} I.,  et~al., 2001, \mn@doi [\aj] {10.1086/323301}, \href
  {http://adsabs.harvard.edu/abs/2001AJ....122.1861S} {122, 1861}

\bibitem[\protect\citeauthoryear{{Sutherland} et~al.,}{{Sutherland}
  et~al.}{2015}]{Sutherland2015}
{Sutherland} W.,  et~al., 2015, \mn@doi [\aap] {10.1051/0004-6361/201424973},
  \href {http://adsabs.harvard.edu/abs/2015A&A...575A..25S} {575, A25}

\bibitem[\protect\citeauthoryear{{Taylor} et~al.,}{{Taylor}
  et~al.}{2011}]{Taylor2011}
{Taylor} E.~N.,  et~al., 2011, \mn@doi [\mnras]
  {10.1111/j.1365-2966.2011.19536.x}, \href
  {http://adsabs.harvard.edu/abs/2011MNRAS.418.1587T} {418, 1587}

\bibitem[\protect\citeauthoryear{{Taylor} et~al.,}{{Taylor}
  et~al.}{2015}]{Taylor2015}
{Taylor} E.~N.,  et~al., 2015, \mn@doi [\mnras] {10.1093/mnras/stu1900}, \href
  {http://adsabs.harvard.edu/abs/2015MNRAS.446.2144T} {446, 2144}

\bibitem[\protect\citeauthoryear{{Thilker} et~al.,}{{Thilker}
  et~al.}{2010}]{Thilker2010}
{Thilker} D.~A.,  et~al., 2010, \mn@doi [\apjl] {10.1088/2041-8205/714/1/L171},
  \href {http://adsabs.harvard.edu/abs/2010ApJ...714L.171T} {714, L171}

\bibitem[\protect\citeauthoryear{{Tonry}, {Blakeslee}, {Ajhar}  \&
  {Dressler}}{{Tonry} et~al.}{2000}]{Tonry2000}
{Tonry} J.~L.,  {Blakeslee} J.~P.,  {Ajhar} E.~A.,   {Dressler} A.,  2000,
  \mn@doi [\apj] {10.1086/308409}, \href
  {http://adsabs.harvard.edu/abs/2000ApJ...530..625T} {530, 625}

\bibitem[\protect\citeauthoryear{{Trayford}, {Theuns}, {Bower}, {Crain},
  {Lagos}, {Schaller}  \& {Schaye}}{{Trayford} et~al.}{2016}]{Trayford2016}
{Trayford} J.~W.,  {Theuns} T.,  {Bower} R.~G.,  {Crain} R.~A.,  {Lagos}
  C.~d.~P.,  {Schaller} M.,   {Schaye} J.,  2016, \mn@doi [\mnras]
  {10.1093/mnras/stw1230}, \href
  {http://adsabs.harvard.edu/abs/2016MNRAS.460.3925T} {460, 3925}

\bibitem[\protect\citeauthoryear{{Tully}, {Mould}  \& {Aaronson}}{{Tully}
  et~al.}{1982}]{Tully1982}
{Tully} R.~B.,  {Mould} J.~R.,   {Aaronson} M.,  1982, \mn@doi [\apj]
  {10.1086/160009}, \href {http://adsabs.harvard.edu/abs/1982ApJ...257..527T}
  {257, 527}

\bibitem[\protect\citeauthoryear{{Warren} et~al.,}{{Warren}
  et~al.}{2007a}]{Warren2007b}
{Warren} S.~J.,  et~al., 2007a, preprint, \href
  {http://adsabs.harvard.edu/abs/2007astro.ph..3037W} {{\!\!}} (\mn@eprint {}
  {astro-ph/0703037})

\bibitem[\protect\citeauthoryear{{Warren} et~al.,}{{Warren}
  et~al.}{2007b}]{Warren2007a}
{Warren} S.~J.,  et~al., 2007b, \mn@doi [\mnras]
  {10.1111/j.1365-2966.2006.11284.x}, \href
  {http://adsabs.harvard.edu/abs/2007MNRAS.375..213W} {375, 213}

\bibitem[\protect\citeauthoryear{{White}, {Navarro}, {Evrard}  \&
  {Frenk}}{{White} et~al.}{1993}]{White1993}
{White} S.~D.~M.,  {Navarro} J.~F.,  {Evrard} A.~E.,   {Frenk} C.~S.,  1993,
  \mn@doi [\nat] {10.1038/366429a0}, \href
  {http://adsabs.harvard.edu/abs/1993Natur.366..429W} {366, 429}

\bibitem[\protect\citeauthoryear{{Willett} et~al.,}{{Willett}
  et~al.}{2013}]{Willett2013}
{Willett} K.~W.,  et~al., 2013, \mn@doi [\mnras] {10.1093/mnras/stt1458}, \href
  {http://adsabs.harvard.edu/abs/2013MNRAS.435.2835W} {435, 2835}

\bibitem[\protect\citeauthoryear{{Williams} et~al.,}{{Williams}
  et~al.}{2016}]{Williams2016}
{Williams} R.~P.,  et~al., 2016, \mn@doi [\mnras] {10.1093/mnras/stw2185},
  \href {http://adsabs.harvard.edu/abs/2016MNRAS.463.2746W} {463, 2746}

\bibitem[\protect\citeauthoryear{{Wyder} et~al.,}{{Wyder}
  et~al.}{2007}]{Wyder2007}
{Wyder} T.~K.,  et~al., 2007, \mn@doi [\apjs] {10.1086/521402}, \href
  {http://adsabs.harvard.edu/abs/2007ApJS..173..293W} {173, 293}

\bibitem[\protect\citeauthoryear{{York} et~al.,}{{York}
  et~al.}{2000}]{York2000}
{York} D.~G.,  et~al., 2000, \mn@doi [\aj] {10.1086/301513}, \href
  {http://adsabs.harvard.edu/abs/2000AJ....120.1579Y} {120, 1579}

\bibitem[\protect\citeauthoryear{{de Jong}, {Verdoes Kleijn}, {Kuijken}  \&
  {Valentijn}}{{de Jong} et~al.}{2013}]{deJong2013}
{de Jong} J.~T.~A.,  {Verdoes Kleijn} G.~A.,  {Kuijken} K.~H.,   {Valentijn}
  E.~A.,  2013, \mn@doi [Experimental Astronomy] {10.1007/s10686-012-9306-1},
  \href {http://adsabs.harvard.edu/abs/2013ExA....35...25D} {35, 25}

\bibitem[\protect\citeauthoryear{{de Jong} et~al.,}{{de Jong}
  et~al.}{2015}]{deJong2015}
{de Jong} J.~T.~A.,  et~al., 2015, \mn@doi [\aap]
  {10.1051/0004-6361/201526601}, \href
  {http://adsabs.harvard.edu/abs/2015A&A...582A..62D} {582, A62}

\bibitem[\protect\citeauthoryear{{de Jong} et~al.,}{{de Jong}
  et~al.}{2017}]{deJong2017}
{de Jong} J.~T.~A.,  et~al., 2017, \mn@doi [\aap]
  {10.1051/0004-6361/201730747}, \href
  {http://adsabs.harvard.edu/abs/2017A&A...604A.134D} {604, A134}

\bibitem[\protect\citeauthoryear{{de Vaucouleurs}}{{de
  Vaucouleurs}}{1959}]{deVaucouleurs1959}
{de Vaucouleurs} G.,  1959, Handbuch der Physik, \href
  {http://adsabs.harvard.edu/abs/1959HDP....53..275D} {53, 275}

\makeatother
\end{thebibliography}

%%%%%%%%%%%%%%%%%%%%%%%%%%%%%%%%%%%%%%%%%%%%%%%%%%

%%%%%%%%%%%%%%%%% APPENDICES %%%%%%%%%%%%%%%%%%%%%

\appendix

\section{Extreme Classification Outliers}
\label{outliers}

\begin{figure*}
	\centering
	\includegraphics[width=\textwidth]{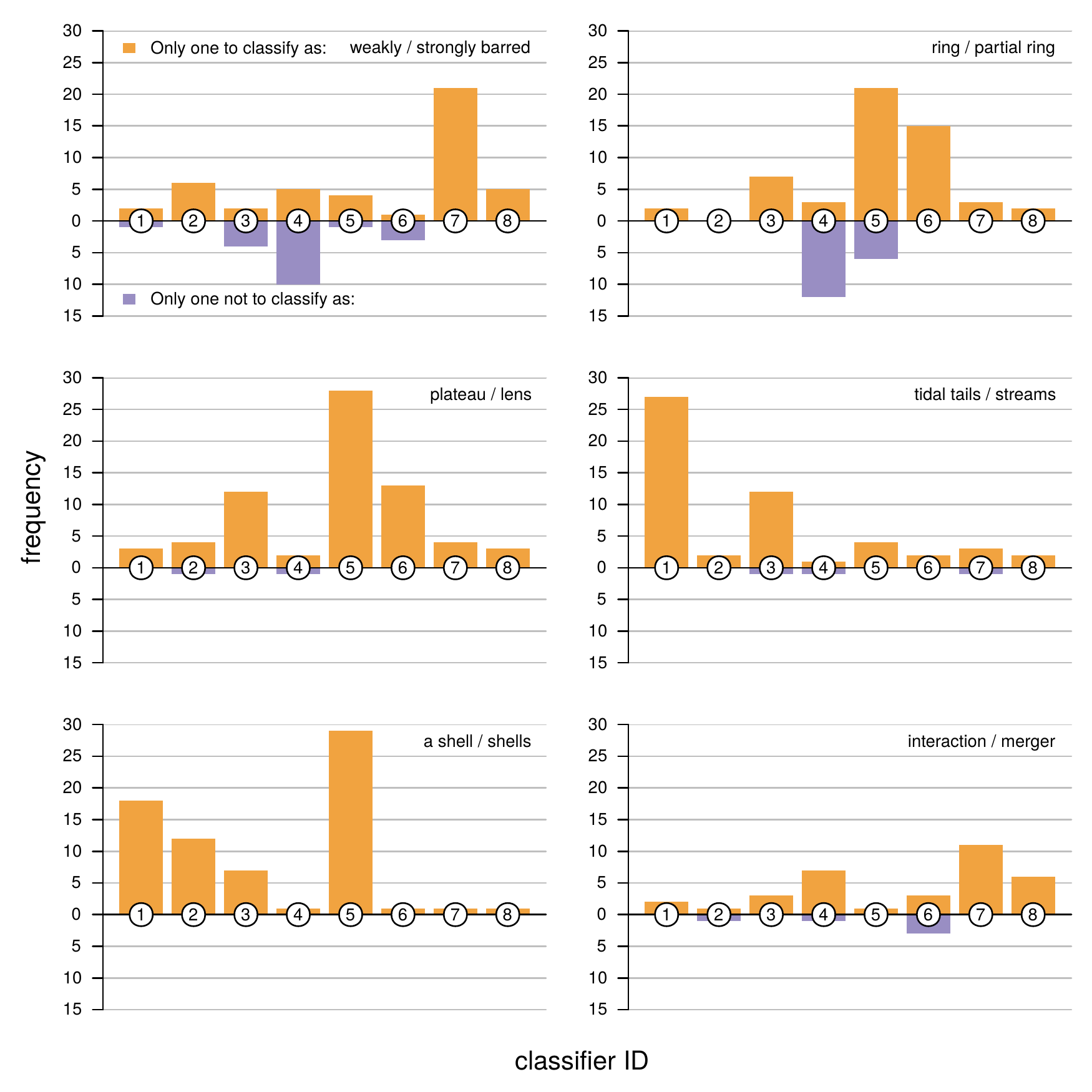}
    \caption{Extreme outliers for classification of the six principal galaxy structures identified in Section \ref{sec:census}, as indicated in the top right of each panel. Classifiers have been randomly assigned a classifier ID in the range $1$ to $8$, arranged along the horizontal axis. Shaded orange bars show how many galaxies out of $472$ a given classifier was the only one to classify in such a way. Shaded purple bars show how many galaxies a given classifier was the only one \textit{not} to classify in such a way.}
    \label{fig:userstats}
\end{figure*}

Combining the classification results from eight independent expert observers allows for the rapid and efficient construction of a robust catalogue of galaxy features for each galaxy in our sample, as evidenced by the high level of concurrence between classifiers for most systems. Figure \ref{fig:userstats} shows extreme outliers by classifier for each of the six principal morphological indicators discussed in Section \ref{sec:census}. Each observer is randomly assigned a classifier ID in the range $1$ to $8$. For each panel, shaded orange bars show how many galaxies out of $472$ a given classifier was the only one to classify in such a way. By contrast, shaded purple bars show how many galaxies a given classifier was the only one \textit{not} to classify in such a way. 

The typical extreme outlier number frequency is of the order $5$ galaxies or less ($\lesssim$$1\%$), never rising above $30$ ($\sim$$6\%$). It's apparent that different classifiers had varying biases towards identifying certain features. For example, classifier $7$ identifies a large number of barred galaxies in systems where no other classifier had identified a bar. Similarly, classifier $5$ has a preference towards identifying ringed, lens and shell type systems, whilst classifier $1$ has a preference towards identifying tidal features. By contrast, classifier $4$ has a preference towards \textit{not} identifying barred and ringed galaxies. Nevertheless, a typical extreme outlier frequency of $\lesssim$$1\%$ indicates a good level of concurrence between classifiers for most systems, underlining the strength of our multi-classifier approach towards galaxy feature identification.

%%%%%%%%%%%%%%%%%%%%%%%%%%%%%%%%%%%%%%%%%%%%%%%%%%

% finish up
\bsp % typesetting comment
\label{lastpage}
\end{document}